\documentclass[12pt]{article}
\usepackage{epsfig}
\usepackage{color}
\usepackage{amssymb,amsmath}
\usepackage{graphicx}
\usepackage{here}
\newcommand{\ba}{\begin{align}}
\newcommand{\ea}{\end{align}}

 \makeatletter
\def\alt{\mathrel{\mathpalette\gl@align<}}
\def\agt{\mathrel{\mathpalette\gl@align>}}
\def\gl@align#1#2{\lower.6ex\vbox{\baselineskip\z@skip\lineskip\z@
\ialign{$\m@th#1\hfil##\hfil$\crcr#2\crcr\sim\crcr}}} \makeatother

\begin{document}
\begin{flushright}
\end{flushright}
\vspace*{1.0cm}

\begin{center}
\baselineskip 20pt 
{\Large\bf 
Prediction on Neutrino Dirac and Majorana Phases \\
and Absolute Mass Scale from the CKM Matrix
}
\vspace{1cm}

{\large 
Naoyuki Haba \ and \ Toshifumi Yamada
} \vspace{.5cm}

{\baselineskip 20pt \it
Graduate School of Science and Engineering, Shimane University, Matsue 690-8504, Japan
}

\vspace{.5cm}

\vspace{1.5cm} {\bf Abstract} \end{center}

In Type-I seesaw model, the lepton flavor mixing matrix (PMNS matrix) and the quark flavor mixing matrix (CKM matrix)
 may be connected implicitly through a relation between the neutrino Dirac Yukawa coupling $Y_D$ and the quark Yukawa couplings. 
In this paper, we study whether $Y_D$ can satisfy, in the flavor basis
 where the charged lepton Yukawa and right-handed neutrino Majorana mass matrices are diagonal,
 the relation $Y_D \propto {\rm diag}(y_d,y_s,y_b)V_{CKM}^T$ or $Y_D \propto {\rm diag}(y_u,y_c,y_t)V_{CKM}^*$
 without contradicting the current experimental data on quarks and neutrino oscillations.
We search for sets of values of the neutrino Dirac CP phase $\delta_{CP}$, Majorana phases $\alpha_2,\alpha_3$,
 and the lightest active neutrino mass that satisfy either of the above relations, with the normal or inverted hierarchy 
 of neutrino mass.
In performing the search, we consider renormalization group evolutions of the quark masses and CKM matrix and the propagation of
 their experimental errors along the evolutions.
We find that only the former relation $Y_D \propto {\rm diag}(y_d,y_s,y_b)V_{CKM}^T$ with the normal neutrino mass hierarchy holds,
 based on which we make a prediction for $\delta_{CP},\,\alpha_2,\,\alpha_3$ and the lightest active neutrino mass.

\thispagestyle{empty}

\newpage

\setcounter{footnote}{0}
\baselineskip 18pt
%

The two flavor mixing matrices, i.e. Pontecorvo-Maki-Nakagawa-Sakata (PMNS) matrix for leptons~\cite{mns,p} and 
 Cabbibo-Kobayashi-Maskawa (CKM) matrix for quarks~\cite{c,km}, are seemingly irrelevant to each other,
 since the former includes two large mixing angles $\theta_{23}\sim45^\circ$ and $\theta_{12}\sim30^\circ$, while the mixing angles of the latter are all below $15^\circ$.
However, if Type-I seesaw mechanism~\cite{seesaw} is operative, there can be a connection between them,
 because the right-handed neutrino Majorana mass matrix that enters into the seesaw mass formula
 distorts the flavor structure of the neutrino Dirac Yukawa coupling,
 so that the active neutrino mass matrix (in the basis where the charged lepton mass is diagonal)
 may have large mixings even when the neutrino Dirac Yukawa coupling (in the same basis) only contains small mixings.

In this paper, we consider the Standard Model (SM) extended with Type-I seesaw mechanism, for which the Yukawa interaction and Majorana mass terms read
\begin{align} 
-{\cal L}&= (Y_u)_{ij} \, \bar{q}_L^i \, i\sigma_2 H^* \, u_R^j + (Y_d)_{ij} \ \bar{q}_L^i \, H \, d_R^j
\nonumber \\
&+(Y_e)_{ij} \ \bar{\ell}_L^i \, H \, e_R^j +(Y_D)_{ij} \ \bar{\ell}_L^i \, i\sigma_2 H^* \, \nu_R^j + \frac{1}{2}(M_N)_{ij} \, \nu_R^i \, \nu_R^j + {\rm H.c.},
\label{lag}
\end{align}
 where $Y_u$, $Y_d$, $Y_e$ and $Y_D$ denote the up-type quark and down-type quark, charged lepton and neutrino Dirac Yukawa couplings, respectively, and $i,j=1,2,3$ are flavor indices.
The mass matrix for active neutrinos, $(m_\nu)_{ij}$, is derived as $m_\nu=-Y_D^* (M_N^*)^{-1}Y_D^\dagger$.
On the other hand, in the flavor basis where the charged lepton mass is diagonal,
 $(m_\nu)_{ij}$ is parametrized in terms of the PMNS matrix, $U_{PMNS}$,
 and the active neutrino masses, $m_1,m_2,m_3$, as 
\begin{align} 
m_\nu&=U_{PMNS}^* \, \begin{pmatrix} 
      m_1 & 0 & 0 \\
      0 & m_2 & 0 \\
      0 & 0 & m_3
   \end{pmatrix}
   U_{PMNS}^\dagger,
\end{align}
 with
\begin{align} 
U_{PMNS}&\equiv   \begin{pmatrix} 
      c_{12}c_{13} & s_{12}c_{13} &  s_{13}e^{-i\,\delta_{CP}}\\
      -s_{12}c_{23}-c_{12}s_{23}s_{13}e^{i\,\delta_{CP}} & c_{12}c_{23}-s_{12}s_{23}s_{13}e^{i\,\delta_{CP}} & s_{23}c_{13}\\
      s_{12}s_{23}-c_{12}c_{23}s_{13}e^{i\,\delta_{CP}} & -c_{12}s_{23}-s_{12}c_{23}s_{13}e^{i\,\delta_{CP}} & c_{23}c_{13}
   \end{pmatrix}
      \begin{pmatrix} 
         1 & 0 & 0 \\
         0 & e^{i\,\alpha_2/2} & 0 \\
         0 & 0 & e^{i\,\alpha_3/2}
      \end{pmatrix}
\\
s_{ab}&\equiv\sin\theta_{ab}, \ \ \ c_{ab}\equiv\cos\theta_{ab} \ \ \ (a,b=1,2,3), \nonumber
\end{align}
 where $\theta_{12},\theta_{23},\theta_{13}$ are the neutrino mixing angles, $\delta_{CP}$ is the Dirac CP phase
 and $\alpha_2,\alpha_3$ are the Majorana CP phases.

We propose the following hypothesis:
In the flavor basis where $Y_e$ and $M_N$ are diagonal,
\begin{align} 
Y_D &= z \, \begin{pmatrix} 
     1  & 0 & 0 \\
      0 & e^{i\phi_2} & 0 \\
      0 & 0 & e^{i\phi_3} \\
   \end{pmatrix} \begin{pmatrix} 
         y_d & 0 & 0\\
         0 & y_s & 0\\
         0 & 0 & y_b\\
      \end{pmatrix} V_{CKM}^T
   \begin{pmatrix} 
     1  & 0 & 0 \\
      0 & e^{i\psi_2} & 0 \\
      0 & 0 & e^{i\psi_3} \\
   \end{pmatrix},
\label{yDyd}\\
{\rm or} \ \ Y_D &= z \, \begin{pmatrix} 
     1  & 0 & 0 \\
      0 & e^{i\phi_2} & 0 \\
      0 & 0 & e^{i\phi_3} \\
   \end{pmatrix} \begin{pmatrix} 
         y_u & 0 & 0\\
         0 & y_c & 0\\
         0 & 0 & y_t\\
      \end{pmatrix} V_{CKM}^*
   \begin{pmatrix} 
     1  & 0 & 0 \\
      0 & e^{i\psi_2} & 0 \\
      0 & 0 & e^{i\psi_3} \\
   \end{pmatrix},
\label{yDyu}
\end{align}
 either of which holds at one renormalization scale $\mu$ located somewhere between the TeV scale
 and the Planck scale.
Here, $y_d,\,y_s,\,y_b,\,y_u,\,y_c,\,y_t$ denote the Yukawa couplings (taken to be real positive) 
 for $d,\,s,\,b,\,u,\,c,\,t$ quarks, respectively.
$z$ is an unspecified complex number and $\phi_2,\phi_3,\psi_2,\psi_3$ are unspecified phases.
We do not center on the different hypotheses in which $V_{CKM}^\dagger$ replaces $V_{CKM}^T$ in Eq.~(\ref{yDyd}),
 or $V_{CKM}$ replaces $V_{CKM}^*$ in Eq.~(\ref{yDyu}).
This is because the combinations $V_{CKM}{\rm diag}(y_d,\,y_s,\,y_b)$ and 
 $V_{CKM}^\dagger{\rm diag}(y_u,\,y_c,\,y_t)$ enter into the Yukawa couplings $Y_d$ and $Y_u$,
 respectively, and the corresponding operators
 $\bar{q}_L\,i\sigma_2H^*\,u_R$ and $\bar{q}_L\,H\,d_R$ have the same chirality as the operator
 $\bar{\ell}_L \, i\sigma_2H^* \, \nu_R$ that is associated with $Y_D$.
This nice feature is spoiled if we take the complex conjugation of one side of Eq.~(\ref{yDyd}) and/or Eq.~(\ref{yDyu}).

The hypothesis Eqs.~(\ref{yDyd},\ref{yDyu}) leads to the following relation between the PMNS and CKM matrices:
If Eq.~(\ref{yDyd}) holds, we obtain
\begin{align}
 &U_{PMNS}^* \, \begin{pmatrix} 
      m_1 & 0 & 0 \\
      0 & m_2 & 0 \\
      0 & 0 & m_3
   \end{pmatrix}
   U_{PMNS}^\dagger
\nonumber \\
   &= -(z^*)^2
      \begin{pmatrix} 
         y_d & 0 & 0 \\
         0 & e^{-i\phi_2}\,y_s & 0 \\
         0 & 0 & e^{-i\phi_3}\,y_b
      \end{pmatrix}
      V_{CKM}^\dagger
      \begin{pmatrix} 
         \frac{1}{M_{N_1}^*} & 0 & 0 \\
         0 & \frac{e^{-2i\psi_2}}{M_{N_2}^*} & 0 \\
         0 & 0 & \frac{e^{-2i\psi_3}}{M_{N_3}^*}
      \end{pmatrix}      
      V_{CKM}^*
     \begin{pmatrix} 
         y_d & 0 & 0 \\
         0 & e^{-i\phi_2}\,y_s & 0 \\
         0 & 0 & e^{-i\phi_3}\,y_b
      \end{pmatrix}\frac{v^2}{2}
\label{pmnsckm},
\end{align}
 where $M_{N_j}$ $(j=1,2,3)$ are the components of the diagonalized Majorana mass matrix $M_N$, and $v\simeq246$~GeV.
If Eq.~(\ref{yDyu}) holds instead, we find
\begin{align}
 &U_{PMNS}^* \, \begin{pmatrix} 
      m_1 & 0 & 0 \\
      0 & m_2 & 0 \\
      0 & 0 & m_3
   \end{pmatrix}
   U_{PMNS}^\dagger
\nonumber \\
   &= -(z^*)^2
      \begin{pmatrix} 
         y_u & 0 & 0 \\
         0 & e^{-i\phi_2}\,y_c & 0 \\
         0 & 0 & e^{-i\phi_3}\,y_t
      \end{pmatrix}
      V_{CKM}
      \begin{pmatrix} 
         \frac{1}{M_{N_1}^*} & 0 & 0 \\
         0 & \frac{e^{-2i\psi_2}}{M_{N_2}^*} & 0 \\
         0 & 0 & \frac{e^{-2i\psi_3}}{M_{N_3}^*}
      \end{pmatrix}      
      V_{CKM}^T
     \begin{pmatrix} 
         y_u & 0 & 0 \\
         0 & e^{-i\phi_2}\,y_c & 0 \\
         0 & 0 & e^{-i\phi_3}\,y_t
      \end{pmatrix}\frac{v^2}{2}
\label{pmnsckm2}.
\end{align}
Experimentally, the Dirac phase $\delta_{CP}$, the Majorana phases $\alpha_2,\alpha_3$ and the absolute scale of the active neutrino mass have not been measured conclusively.
Eq.~(\ref{pmnsckm}) or (\ref{pmnsckm2}) hence contains 12 real undetermined variables, which are
\begin{align} 
\delta_{CP}, \ \alpha_2, \ \alpha_3, \ \phi_2, \ \phi_3, \ m_1, \ \frac{z^2}{M_{N_1}} \, ({\rm complex}),
\ \frac{z^2\, e^{2i\psi_2}}{M_{N_2}} \, ({\rm complex}), \ \frac{z^2\, e^{2i\psi_3}}{M_{N_3}} \, ({\rm complex}),
\end{align}
 where it should be noted that $z$, $M_{N_i}$ and $\psi_i$ appear only in the above combination.
On the other hand, Eq.~(\ref{pmnsckm}) or (\ref{pmnsckm2}) yields 6 complex equations, since both sides are complex symmetric matrices.
Therefore, Eq.~(\ref{pmnsckm}) or (\ref{pmnsckm2}) can, in principle, fix the 12 undetermined variables.
As a matter of fact, some undetermined variables are phases and hence it is highly non-trivial that the solution to Eq.~(\ref{pmnsckm}) or (\ref{pmnsckm2}) exists.
In the rest of paper, we study whether the solution to Eq.~(\ref{pmnsckm}) or (\ref{pmnsckm2}) exists
 for the normal ($m_3>m_2>m_1$) and inverted hierarchy ($m_2>m_1>m_3$) of the active neutrino mass,
 and if it does, we draw a prediction for $\delta_{CP}$, $\alpha_2,\alpha_3$ and the lightest active neutrino mass.
We pay attention to the fact that some of the quantities that enter into Eqs.~(\ref{pmnsckm},\ref{pmnsckm2})
 are subject to sizable experimental errors, which causes ambiguity in the solution.
We also note that different solutions may be obtained depending on the scale at which Eq.~(\ref{yDyd}) or (\ref{yDyu}) holds,
 due to renormalization group (RG) evolutions of the quark Yukawa couplings and CKM matrix.
Therefore, we scrutinize their RG evolutions and
 how the experimental errors of the quark masses, mixing angles and Kobayashi-Maskawa phase propagate along the evolutions.
In contrast, we directly use the values of neutrino mixing angles and mass differences measured in neutrino oscillation experiments;
 this is justified by assuming that all the components of $Y_D$ are much below 1, or equivalently $\vert z\vert\ll1$,
 so that terms like $Y_D Y_D^\dagger Y_D$ in RG equations are negligible and the RG evolutions change $Y_D$ only by an overall constant that can be absorbed into the number $z$ in Eqs.~(\ref{yDyd},\ref{yDyu}).

Our calculation of RG evolutions of quark Yukawa couplings and $V_{CKM}$ proceeds by the following steps.
All the renormalization scales are in $\overline{MS}$ scheme.
\\

\noindent
(I) Below a scale $\mu_{EW}\sim M_Z$, we work in 5 or 4-flavor QCD$\times$QED theory
 (decoupling of $b$ at $\mu=m_b(m_b)$ is properly taken into account~\cite{decoupling}).
We solve QCD 3-loop and QED 1-loop RG equation~\cite{qcdqedrge} for QCD coupling $\alpha_s$ in the range $\mu_{EW}>\mu >2$~GeV
  and that for QED coupling $\alpha_{em}$ in the range $\mu_{EW}>\mu >13$~GeV
 (we ignore QED effects below 13~GeV),
 with the initial values of $\alpha_s^{(5)}(M_Z)=0.1182$ and $\alpha(M_Z)=1/127.950$ quoted from the Particle Data Group~\cite{pdg}.
\\

\noindent
(II) We solve QCD 3-loop and QED 1-loop RG equation~\cite{massrge} for $u,d,s,c,b$ quark masses up to $\mu=\mu_{EW}$.
 For average $u$-$d$ mass and $s$ mass, we quote the results of lattice calculations~\cite{lattice,latticereview}, 
 $\frac{1}{2}(m_u+m_d)(2~{\rm GeV})=3.373(80)~{\rm MeV}$ and $m_s(2~{\rm GeV})=92.0(2.1)~{\rm MeV}$.
 For $u$-$d$ mass ratio, we refer to an estimate in Ref.~\cite{latticereview}, $m_u/m_d=0.46(3)$.
 For $c$ and $b$ masses, we adopt the results of QCD sum rule analyses~\cite{cb},
  $m_c(3~{\rm GeV})=0.986-9(\alpha_s^{(5)}(M_Z)-0.1189)/0.002\pm0.010~{\rm GeV}$
  and $m_b(m_b)=4.163+7(\alpha_s^{(5)}(M_Z)-0.1189)/0.002\pm0.014~{\rm GeV}$.
 \\
 
\noindent
(III) We match 5-flavor QCD$\times$QED theory with the full SM at $\mu=\mu_{EW}$.
 For $t$ quark mass, we adopt the pole mass obtained from the exclusive $t$ pair production cross section at the LHC~\cite{toppole},
 $M_t =173.7^{+2.3}_{-2.1}~{\rm GeV}$, and for $W,Z$ and Higgs boson masses and $G_F$, we use Particle Data Group values~\cite{pdg}.
We evaluate QCD 2-loop threshold corrections of $t$ quark on $\alpha_s$~\cite{decoupling}
  and QED 1-loop threshold corrections of $t$ quark and $W$ boson on $\alpha_{em}$~\cite{fanchiotti} 
  to obtain QCD and QED gauge couplings in the SM.
We employ the results of Refs.~\cite{Jegerlehner:2001fb,Jegerlehner:2003py,Bezrukov:2012sa,topthreshold,threshold} implemented in the code~\cite{code}
  to compute the $t$ quark Yukawa coupling $y_t(\mu_{EW})$ with QCD 4-loop and QED 2-loop threshold corrections, 
  and to compute the Higgs quartic coupling $\lambda_H(\mu_{EW})$, running Higgs vacuum expectation value (VEV) $v(\mu_{EW})$ and running weak mixing angle $\sin^2\theta_W(\mu_{EW})$ with QED 2-loop and QCD 1-loop threshold corrections.
 We reconstruct the CKM matrix $\bar{V}_{CKM}$ from up-to-date values of the Wolfenstein parameters reported by the CKMfitter~\cite{ckmfitter},
 $A=0.8250^{+0.0071}_{-0.0111}$, $\lambda=0.22509^{+0.00029}_{-0.00028}$, $\bar{\rho}=0.1598^{+0.0076}_{-0.0072}$ and $\bar{\eta}=0.3499^{+0.0063}_{-0.0061}$.
Finally, we derive the running Yukawa matrices for quarks by neglecting threshold corrections for the CKM matrix as
\begin{align} 
y_i(\mu_{EW})&=\sqrt{2}m_i(\mu_{EW})/v(\mu_{EW}) \ \ \ \ \ (i=u,d,s,c,b),
\nonumber \\
Y_d(\mu_{EW})&=\bar{V}_{CKM}  
\begin{pmatrix} 
      y_d(\mu_{EW}) & 0 & 0 \\
      0 & y_s(\mu_{EW}) & 0 \\
      0 & 0 & y_b(\mu_{EW}) \\
   \end{pmatrix},
\nonumber \\
Y_u(\mu_{EW})&=   
\begin{pmatrix} 
      y_u(\mu_{EW}) & 0 & 0 \\
      0 & y_c(\mu_{EW}) & 0 \\
      0 & 0 & y_t(\mu_{EW}) \\
   \end{pmatrix}.
\nonumber \\
\end{align}
Although insignificant in our analysis, we further derive the running Yukawa matrix for charged leptons
 from the Particle Data Group values of lepton masses,
 by adding 1-loop threshold corrections and then dividing them by the running Higgs VEV $v(\mu_{EW})$.
\\

\noindent
(IV) We solve the full 3-loop RG equations of the SM~\cite{smrge} in the range $\mu_{EW}\leq\mu\leq10^{18}$~GeV.
\\

\noindent
(V) At various scales $\mu$, we derive the running Yukawa couplings (taken to be real positive), $y_i(\mu)$ $(i=u,c,t,d,s,b)$,
 and the running CKM matrix, $V_{CKM}(\mu)$, in the following manner.
We diagonalize the Yukawa matrices at scale $\mu$ as
\begin{align} 
V_{uL}(\mu)Y_u(\mu)Y^\dagger_u(\mu)V_{uL}^\dagger(\mu)&=
\begin{pmatrix} 
      y_u(\mu)^2 & 0 & 0 \\
      0 & y_c(\mu)^2 & 0 \\
      0 & 0 & y_t(\mu)^2 \\
   \end{pmatrix},
\nonumber \\
V_{dL}(\mu)Y_d(\mu)Y^\dagger_d(\mu)V_{dL}^\dagger(\mu)&=
\begin{pmatrix} 
      y_d(\mu)^2 & 0 & 0 \\
      0 & y_s(\mu)^2 & 0 \\
      0 & 0 & y_b(\mu)^2 \\
   \end{pmatrix},
\end{align}
where $V_{uL}(\mu),V_{dL}(\mu)$ are unitary matrices depending on $\mu$.
Then we calculate $V_{CKM}(\mu)=V_{uL}(\mu)V_{dL}^\dagger(\mu)$,
 and further decompose it into physical three mixing angles, $\theta_{ij}^{ckm}(\mu)$, and one CP phase, $\delta_{km}(\mu)$, as
\begin{align} 
\sin\theta_{13}^{ckm}(\mu)&=|V_{ub}(\mu)|, \ \ \ \sin\theta_{12}^{ckm}(\mu)=\frac{|V_{us}(\mu)|}{\sqrt{1-|V_{ub}(\mu)|^2}}, \ \ \
\sin\theta_{23}^{ckm}(\mu)=\frac{|V_{cb}(\mu)|}{\sqrt{1-|V_{ub}(\mu)|^2}},
\nonumber \\
e^{i\,\delta_{km}(\mu)}&=\left.
\left(\frac{V_{us}V_{cb}V_{ub}^*V_{cs}^*}{|V_{ub}||V_{us}||V_{cb}|}+
\frac{|V_{ub}||V_{us}||V_{cb}|}{1-|V_{ub}|^2}\right)\frac{1-|V_{ub}|^2}{\sqrt{1-|V_{ub}|^2-|V_{us}|^2}\sqrt{1-|V_{ub}|^2-|V_{cb}|^2}}\right\vert_\mu.
\label{decompose}
\end{align}

We estimate uncertainties of the Yukawa couplings and CKM matrix at each scale $\mu$ as follows:
\begin{itemize}

\item For each running Yukawa coupling $y_i(\mu)$ $(i=u,c,t,d,s,b)$, we consider the propagation of the experimental error of its corresponding mass \textit{only} and estimate its uncertainty, $\Delta y_i(\mu)$, as
\begin{align} 
(\Delta y_i(\mu))^2&=
\left(\frac{\partial y_i(\mu)}{\partial (m_u+m_d)(2~{\rm GeV})}\right)^2\Delta(m_u+m_d)(2~{\rm GeV})^2
+\left(\frac{\partial y_i(\mu)}{\partial m_u/m_d}\right)^2\Delta(m_u/m_d)^2 
\nonumber \\ &(i=u,d),
\nonumber \\
\Delta y_s(\mu)&=\left\vert \frac{\partial y_s(\mu)}{\partial m_s(2~{\rm GeV})}\right\vert\Delta m_s(2~{\rm GeV}), \ \ 
\Delta y_c(\mu)=\left\vert \frac{\partial y_c(\mu)}{\partial m_c(3~{\rm GeV})}\right\vert\Delta m_c(3~{\rm GeV}),
\nonumber \\
\Delta y_b(\mu)&=\left\vert \frac{\partial y_b(\mu)}{\partial m_b(m_b)}\right\vert\Delta m_b(m_b), \ \
\Delta y_t(\mu)=\left\vert \frac{\partial y_t(\mu)}{\partial M_t}\right\vert\Delta M_t,
\end{align}
 where we take $\Delta M_t=2.3$~GeV.

\item For the running CKM mixing angles $\theta_{ij}^{ckm}(\mu)$ and CP phase $\delta_{km}(\mu)$,
  we estimate their uncertainties, $\Delta \theta_{ij}(\mu)$ and $\Delta \delta_{km}(\mu)$, by assuming that experimental errors of the Wolfenstein parameters are maximally correlated, and thereby linearly adding errors propagating from these experimental errors as
\begin{align} 
\Delta \theta_{ij}(\mu)&=\sum_{Q=A,\lambda,\bar{\rho},\bar{\eta}}\left\vert\frac{\partial \theta_{ij}^{ckm}(\mu)}{\partial Q}\right\vert\Delta Q, \ \ \ \ \
\Delta \delta_{km}(\mu)=\sum_{Q=A,\lambda,\bar{\rho},\bar{\eta}}\left\vert\frac{\partial \delta_{km}(\mu)}{\partial Q}\right\vert\Delta Q,
\label{wolferror}
\end{align}
 where we take $\Delta A=0.0111$, $\Delta \lambda=0.00029$, $\bar{\rho}=0.0076$ and $\bar{\eta}=0.0063$.
\end{itemize}

In Table~\ref{running}, we present the running quark Yukawa couplings, CKM mixing angles and phase, and their uncertainties
 at scales $\mu=10^4,~10^6,~10^{12},~10^{15},~10^{18}$~GeV in $\overline{MS}$ scheme,
 evaluated by taking the matching scale as $\mu_{EW}=M_Z$.
\begin{table}[H]
\begin{center}
  \caption{Running Yukawa couplings, CKM mixing angles and phase, and their uncertainties at various scales $\mu$ in $\overline{MS}$ scheme, evaluated by taking the matching scale as $\mu_{EW}=M_Z$.
   For each Yukawa coupling $y_i$, the uncertainty is estimated by considering the propagation of the experimental error of its corresponding mass only (for $u$ and $d$ quarks, we consider both errors of $(m_u+m_d)(2~{\rm GeV})$ and $m_u/m_d$).
  For the CKM mixing angles and phase, we assume maximal correlation among the errors of the Wolfenstein parameters,
   thereby linearly adding uncertainties that propagate from the experimental errors, as in Eq.~(\ref{wolferror}).
  }
  \begin{tabular}{|c|ccccc|} \hline
    scale $\mu$ & $10^4$~GeV  & $10^6$~GeV & $10^{12}$~GeV & $10^{15}$~GeV & $10^{18}$~GeV\\ \hline
    $10^6~y_u$           &5.43(27)&4.56(23)&3.22(16)&2.83(14)&2.53(13)\\
    $10^3~y_c$           &2.794(28)&2.347(24)&1.657(17)&1.457(15)&1.301(13)\\
    $y_t$                       &0.777(13)&0.668(13)&0.493(13)&0.440(12)&0.398(12)\\ \hline
    $10^6~y_d$           &11.89(37)&10.02(31)&7.17(22)&6.36(20)&5.72(18)\\
    $10^3~y_s$           &0.2366(54)&0.1995(46)&0.1428(33)&0.1265(29)&0.1140(26)\\
    $y_b$                              &0.012195(43)&0.010042(35)&0.006879(24)&0.006008(21)&0.005349(19)\\ \hline
    $\theta^{ckm}_{12}$~(rad)             &0.22705(30)&0.22705(30)&0.22707(30)&0.22707(30)&0.22707(30)\\
    $10^3~\theta^{ckm}_{13}$~(rad)  &3.85(16)  &3.94(16)  &4.12(17)   &4.18(17)  &4.22(17)\\
    $10^2~\theta^{ckm}_{23}$~(rad)  &4.332(69)&4.436(71)&4.635(74)&4.703(75)&4.758(76)\\
    $\delta_{km}/(2\pi)$           &0.1819(39)&0.1819(39)&0.1819(39)&0.1819(39)&0.1819(39)\\ \hline
  \end{tabular}
  \label{running}
  \end{center}
\end{table}
\noindent
In Table~\ref{running2}, we present the results obtained by taking $\mu_{EW}=160$~GeV instead.
The sizable discrepancy in the Yukawa couplings for $\mu_{EW}=M_Z$ and $\mu_{EW}=160$~GeV
 is due to keen $\mu$-dependence of the running Higgs VEV $v(\mu)$, which originates from
 tadpole contributions to $v(\mu)$~\cite{tadpole} that are enhanced by $(N_c M_t^4/(M_W^2M_h^2))^l$
 as the number of loops $l$ increases.
\begin{table}[H]
\begin{center}
  \caption{Same as Table~\ref{running}, except that we take the matching scale as $\mu_{EW}=160$~GeV.}
  \begin{tabular}{|c|ccccc|} \hline
    scale $\mu$ & $10^4$~GeV  & $10^6$~GeV & $10^{12}$~GeV & $10^{15}$~GeV & $10^{18}$~GeV\\ \hline
    $10^6~y_u$           &5.11(26)&4.29(22)&3.03(15)&2.67(13)&2.38(12)\\
    $10^3~y_c$           &2.626(27)&2.206(22)&1.558(16)&1.370(14)&1.224(12)\\
    $y_t$                       &0.778(13)&0.669(13)&0.494(13)&0.441(13)&0.399(12)\\ \hline
    $10^6~y_d$           &11.17(35)&9.42(30)&6.75(21)&5.98(19)&5.39(17)\\
    $10^3~y_s$           &0.2225(51)&0.1876(43)&0.1343(31)&0.1190(27)&0.1072(24)\\
    $y_b$                              &0.011525(37)&0.009491(31)&0.006504(21)&0.005681(18)&0.005058(16)\\ \hline
    $\theta^{ckm}_{12}$~(rad)             &0.22704(30)&0.22705(30)&0.22706(30)&0.22707(30)&0.22707(30)\\
    $10^3~\theta^{ckm}_{13}$~(rad)  &3.83(16)  &3.92(16)  &4.09(17)   &4.15(17)  &4.20(17)\\
    $10^2~\theta^{ckm}_{23}$~(rad)  &4.310(69)&4.413(71)&4.612(74)&4.680(75)&4.735(76)\\
    $\delta_{km}/(2\pi)$           &0.1819(39)&0.1819(39)&0.1819(39)&0.1819(39)&0.1819(39)\\ \hline
  \end{tabular}
  \label{running2}
  \end{center}
\end{table}

For the neutrino mixing angles and mass differences, we employ Particle Data Group values~\cite{pdg}:
 $\sin^2\theta_{12}=0.307\pm0.013$,  $\sin^2\theta_{13}=0.0210\pm0.0011$,
  $\Delta m^{2}_{21}=(7.53\pm0.18)\times10^{-5}$~eV$^2$, 
  $\sin^2\theta_{23}=0.51\pm0.04$ (normal hierarchy), $\sin^2\theta_{23}=0.50\pm0.04$ (inverted hierarchy), 
  $\vert \Delta m^{2}_{32} \vert=(2.45\pm0.05)\times10^{-3}$~eV$^2$ (normal hierarchy),
  $\vert \Delta m^{2}_{32} \vert=(2.52\pm0.05)\times10^{-3}$~eV$^2$ (inverted hierarchy).
We neglect RG evolutions of $Y_D$ and directly use the above experimental values in our analysis.

Using the running quark Yukawa couplings, running CKM matrix, neutrino mass differences and mixing angles obtained above,
 we search for the solution to Eq.~(\ref{pmnsckm}) or (\ref{pmnsckm2}),
 and further estimate its uncertainty that stems from experimental errors of the quark masses, Wolfenstein parameters,
 and neutrino oscillation parameters.
We test four cases,

\noindent
(A) Eq.~(\ref{yDyd}) holds and the neutrino mass hierarchy is normal;

\noindent
(B) Eq.~(\ref{yDyd}) holds and the neutrino mass hierarchy is inverted;

\noindent
(C) Eq.~(\ref{yDyu}) holds and the neutrino mass hierarchy is normal;

\noindent
(D) Eq.~(\ref{yDyu}) holds and the neutrino mass hierarchy is inverted.

\noindent
For Case(A), we conduct the solution search by reformulating Eq.~(\ref{pmnsckm}) into the equation below,
\begin{align}
 &\begin{pmatrix} 
         1/y_d & 0 & 0 \\
         0 & 1/y_s & 0 \\
         0 & 0 & 1/y_b
      \end{pmatrix}
      U_{PMNS}^* \begin{pmatrix} 
      m_1 & 0 & 0 \\
      0 & m_2 & 0 \\
      0 & 0 & m_3
   \end{pmatrix}
   U_{PMNS}^\dagger\begin{pmatrix} 
         1/y_d^2 & 0 & 0 \\
         0 & 1/y_s^2 & 0 \\
         0 & 0 & 1/y_b^2
      \end{pmatrix}
     U_{PMNS}
\nonumber \\
&\times     \begin{pmatrix} 
      m_1 & 0 & 0 \\
      0 & m_2 & 0 \\
      0 & 0 & m_3
   \end{pmatrix}
   U_{PMNS}^T
   \begin{pmatrix} 
         1/y_d & 0 & 0 \\
         0 & 1/y_s & 0 \\
         0 & 0 & 1/y_b
      \end{pmatrix}
\nonumber \\
   &= \vert z\vert^2  \frac{v^4}{4} \begin{pmatrix} 
         1 & 0 & 0 \\
         0 & e^{-i\phi_2}& 0 \\
         0 & 0 & e^{-i\phi_3}
      \end{pmatrix}
      V_{CKM}^\dagger
      \begin{pmatrix} 
         \frac{1}{M_{N_1}^2} & 0 & 0 \\
         0 & \frac{1}{M_{N_2}^2} & 0 \\
         0 & 0 & \frac{1}{M_{N_3}^2}
      \end{pmatrix}      
      V_{CKM}
     \begin{pmatrix} 
         1 & 0 & 0 \\
         0 & e^{i\phi_2} & 0 \\
         0 & 0 & e^{i\phi_3}
      \end{pmatrix},
\label{pmnsckmalt}
\end{align}
 and then taking the following steps (the procedures are analogous for Cases(B),(C),(D)):
 \\
 
\noindent
(i) We select a set of 'input values' of $y_d,y_s,y_b$ and $\theta_{12},\theta_{13},\theta_{23},\vert\Delta m_{32}^2\vert,\Delta m_{21}^2$ from the 2$\sigma$ range.
For $y_d,y_s,y_b$, we quote the values at $\mu=10^{18}$~GeV evaluated by taking $\mu_{EW}=M_Z$ in Table~\ref{running}.
Namely, we take
\begin{align} 
y_d&\in [5.36,~6.08]\times10^{-6}, \ \ \ y_s\in [1.088,~1.192]\times10^{-4}, \ \ \ y_b\in [5.293,~5.405]\times10^{-3},
\nonumber \\
\sin^2\theta_{12}&\in [0.281,~0.333], \ \ \ \sin^2\theta_{13}\in [0.0188,~0.0232], \ \ \
\sin^2\theta_{23}\in [0.43,~0.59],
\nonumber \\
\Delta m_{21}^2&\in [7.17,~7.89]\times10^{-5}~{\rm eV}^2, \ \ \
\vert\Delta m_{32}^2\vert\in [2.35,~2.55]\times10^{-3}~{\rm eV}^2.
\label{inputsrange}
\end{align}
\\ 

\noindent
(ii) We randomly generate a set of 'trial values' of the neutrino Dirac CP phase and Majorana CP phases and the logarithm of the lightest neutrino mass, $(\delta_{CP},\,\alpha_2,\,\alpha_3,\,\log(m_1))$, which vary in the following range:
\begin{align} 
2\pi&>\delta_{CP}\geq0, \ \ \ 2\pi>\alpha_2\geq0, \ \ \ 2\pi>\alpha_3\geq0, \ \ \
-1>\log_{10}(m_1/{\rm eV})>-3,
\label{randomrange}
\end{align}
 where the maximum of $m_1$ corresponds to the quasi-degenerate case with $m_1\simeq m_2\simeq m_3$,
 and the minimum corresponds to the case in which the lightest neutrino mass is negligible $m_1\ll m_2$.
(For Cases(B),(D), $m_1$ should be replaced with $m_3$.)
\\

\noindent
(iii) We insert the above 'input values' and 'trial values' into the left-hand side of Eq.~(\ref{pmnsckmalt}), numerically diagonalize the left-hand side with a unitary matrix $V_{test}$, 
 and decompose $V_{test}$ into three mixing angles $\theta_{12,test}^{ckm},\theta_{13,test}^{ckm},\theta_{23,test}^{ckm}$
 and one CP phase $\delta_{km,test}$ in a way analogous to Eq.~(\ref{decompose}).
When decomposing $V_{test}$, we rearrange the rows of $V_{test}$ so that $\theta_{12,test}>\theta_{23,test}>\theta_{13,test}^{ckm}$ holds,
 which is justifiable because the ordering of $M_{N_1},M_{N_2},M_{N_3}$ is arbitrary.
We study if the mixing angles $\theta_{12,test}^{ckm},\theta_{13,test}^{ckm},\theta_{23,test}^{ckm}$ and CP phase $\delta_{km,test}$ fit within the 2$\sigma$ range of $\theta_{12}^{ckm},\theta_{13}^{ckm},\theta_{23}^{ckm},\delta_{km}$ at $\mu=10^{18}$~GeV evaluated with $\mu_{EW}=M_Z$ (Table~\ref{running}), namely, we check
\begin{align} 
0.22767&\geq\theta_{12,test}^{ckm}\geq0.22647, \ \ \
4.59\times10^{-3}\geq\theta_{13,test}^{ckm}\geq3.88\times10^{-3},
\nonumber \\
4.910\times10^{-2}&\geq\theta_{23,test}^{ckm}\geq4.606\times10^{-2}, \ \ \
0.1897\geq\delta_{km,test}\geq0.1741.
\label{ckmselection}
 \end{align}
If the above inequalities all hold, the corresponding set of 'trial values' $(\delta_{CP},~\alpha_2,~\alpha_3,~\log(m_1))$
 is regarded as a solution to Eq.~(\ref{pmnsckm}).

The results are as follows.
\begin{itemize}
\item In Cases(B),(C),(D),
 we have generated $1.6\times10^{10}$ random sets of 'trial values' of
 
\noindent
 $(\delta_{CP},\,\alpha_2,\,\alpha_3,\,\log(m_1))$
 for each set of 'input values' of $y_d,y_s,y_b$ and $\theta_{12},\theta_{13},\theta_{23},\vert\Delta m_{32}^2\vert,\Delta m_{21}^2$,
 and found no solution to Eq.~(\ref{pmnsckm}) or (\ref{pmnsckm2}).
There is no solution even when we loosen the criteria of (iii) and allow 
 $\theta_{12,test}^{ckm},\theta_{13,test}^{ckm},\theta_{23,test}^{ckm},\delta_{km,test}$ to fit within 
 the 5$\sigma$ range.
 
 \item In Case(A), we have generated $1.6\times10^{10}$ random sets of 'trial values' of 

\noindent
 $(\delta_{CP},\,\alpha_2,\,\alpha_3,\,\log(m_1))$ for each set of 'input values',
 and found solutions to Eq.~(\ref{pmnsckm}) for $\sin^2\theta_{23}=0.43,\,0.47$,
 whereas no solution is found for $\sin^2\theta_{23}=0.51,\,0.55,\,0.59$.
The value of $\delta_{CP}$ in the solutions exhibits a correlation with $\theta_{13,test}^{ckm}$,
 so we plot the solutions on the plane of $(\theta_{13,test}^{ckm},\,\delta_{CP})$.
Since the values of $\alpha_2,\alpha_3,m_1$ in the solutions are strongly correlated with $\delta_{CP}$,
 we further plot the solutions on the planes of $(\delta_{CP},\,\alpha_2)$, $(\delta_{CP},\,\alpha_3)$ and
 $(\delta_{CP},\,m_1)$.
Additionally, we calculate, for individual solutions, $m_{ee}$, the quantity measured in neutrinoless double $\beta$-decay experiments, as
 \begin{align} 
m_{ee}&=\left\vert\left[ U_{PMNS}^* \begin{pmatrix} 
      m_1 & 0 & 0 \\
      0 & m_2 & 0 \\
      0 & 0 & m_3
   \end{pmatrix}
   U_{PMNS}^\dagger \right]_{ee}\right\vert,
 \end{align}
  and plot the solutions on the plane of $(\delta_{CP},\,m_{ee})$.
The results are displayed in Figures~\ref{043047},\,\ref{s12pm},\,\ref{s13pm},\,\ref{ydpm},\,\ref{yspm},\,\ref{ybpm},
 whose corresponding 'input values' are listed in Table~\ref{inputs}.
Note that the scattering of dots in each figure represents uncertainty of the solution
 due to the uncertainties of $\theta_{12}^{ckm},\theta_{13}^{ckm},\theta_{23}^{ckm},\delta_{km}$.
\begin{table}[H]
\begin{center}
  \caption{Input values of $y_d,y_s,y_b$ and $\theta_{12},\theta_{13},\theta_{23},\vert\Delta m_{32}^2\vert,\Delta m_{21}^2$
  which correspond to Figures~\ref{043047}--\ref{ybpm}. }
  \begin{tabular}{|c|cccccc|} \hline
    Input values & Figure~\ref{043047} & Figure~\ref{s12pm}  & Figure~\ref{s13pm} & Figure~\ref{ydpm} & Figure~\ref{yspm} & Figure~\ref{ybpm}\\ \hline
    $10^6~y_d$      &5.72&5.72&5.72&5.36{\rm or}6.08&5.72&5.72\\
    $10^4~y_s$      &1.140&1.140&1.140&1.140&1.088{\rm or}1.192&1.140\\
    $10^3~y_b$      &5.349&5.349&5.349&5.349&5.349&5.293{\rm or}5.405\\ \hline
    $\sin^2\theta_{12}$  &0.307&0.281{\rm or}0.333&0.307&0.307&0.307&0.307\\
    $10^2~\sin^2\theta_{13}$  &2.10&2.10&1.88{\rm or}2.32&2.10&2.10&2.10\\
    $\sin^2\theta_{23}$  &0.43{\rm or}0.47&0.47&0.47&0.47&0.47&0.47\\ \hline
    $10^5~\Delta m_{21}^2/$eV$^2$     &7.53&7.53&7.53&7.53&7.53&7.53\\
    $10^3~\vert\Delta m_{32}^2\vert/$eV$^2$     &2.45&2.45&2.45&2.45&2.45&2.45\\ \hline
  \end{tabular}
  \label{inputs}
  \end{center}
\end{table}
\begin{figure}[H]
\begin{center}
\includegraphics[width=80mm]{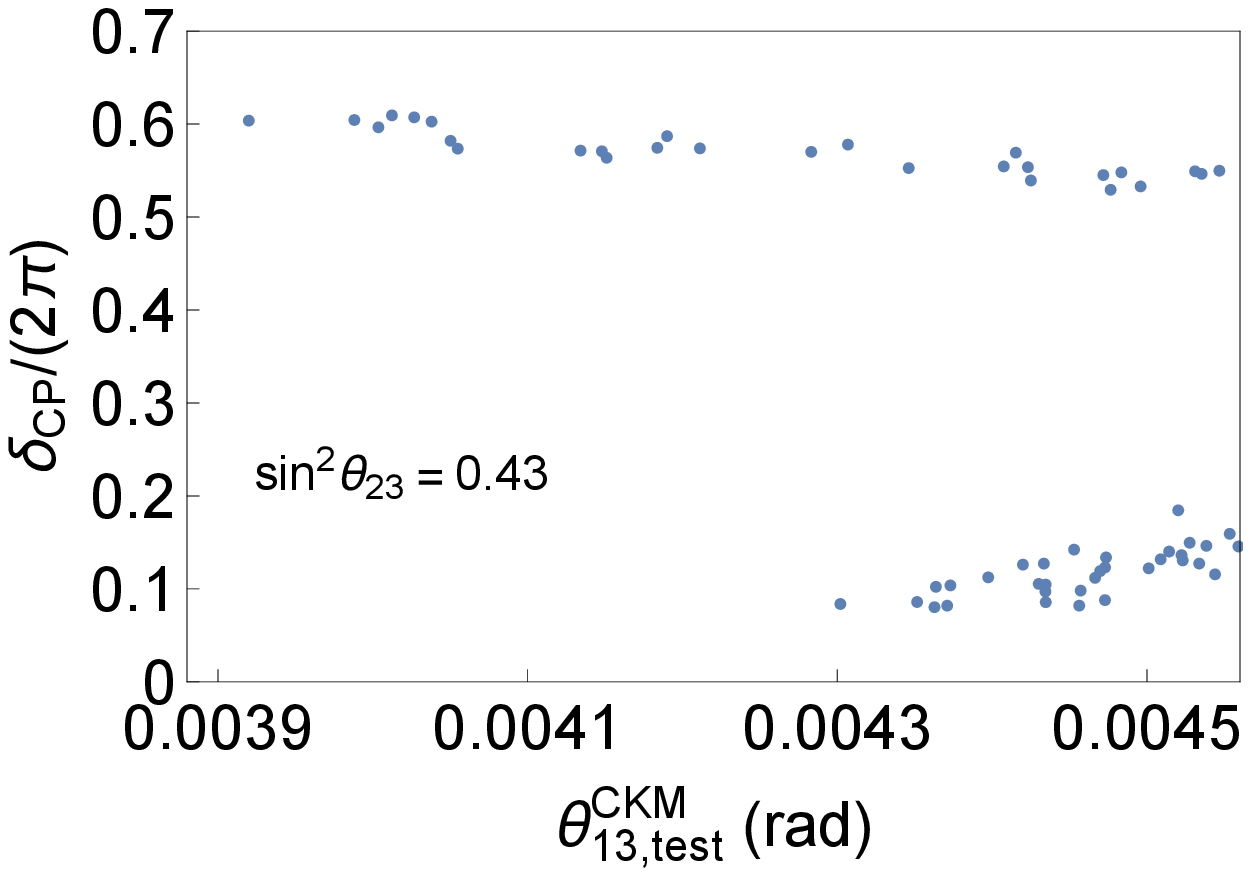}
\includegraphics[width=80mm]{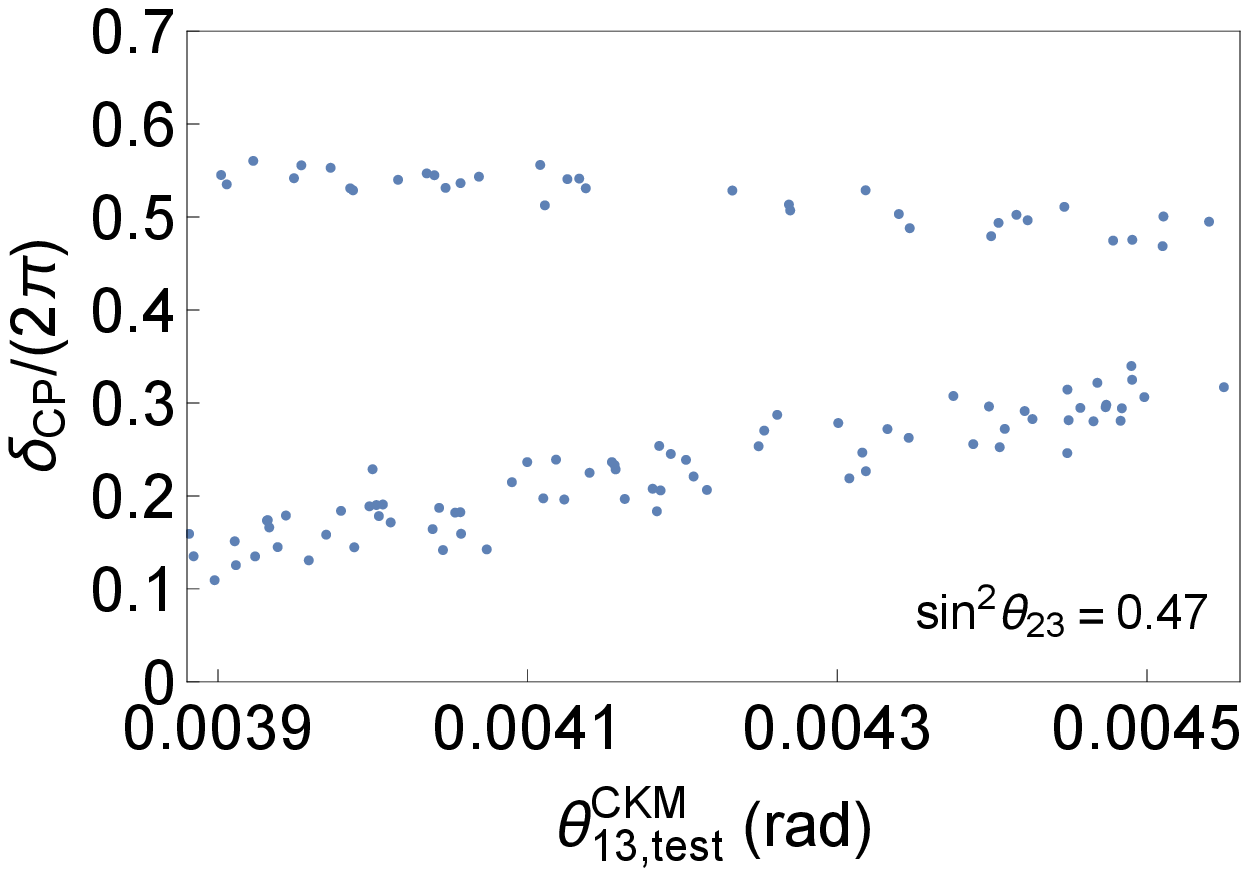}\\
\includegraphics[width=80mm]{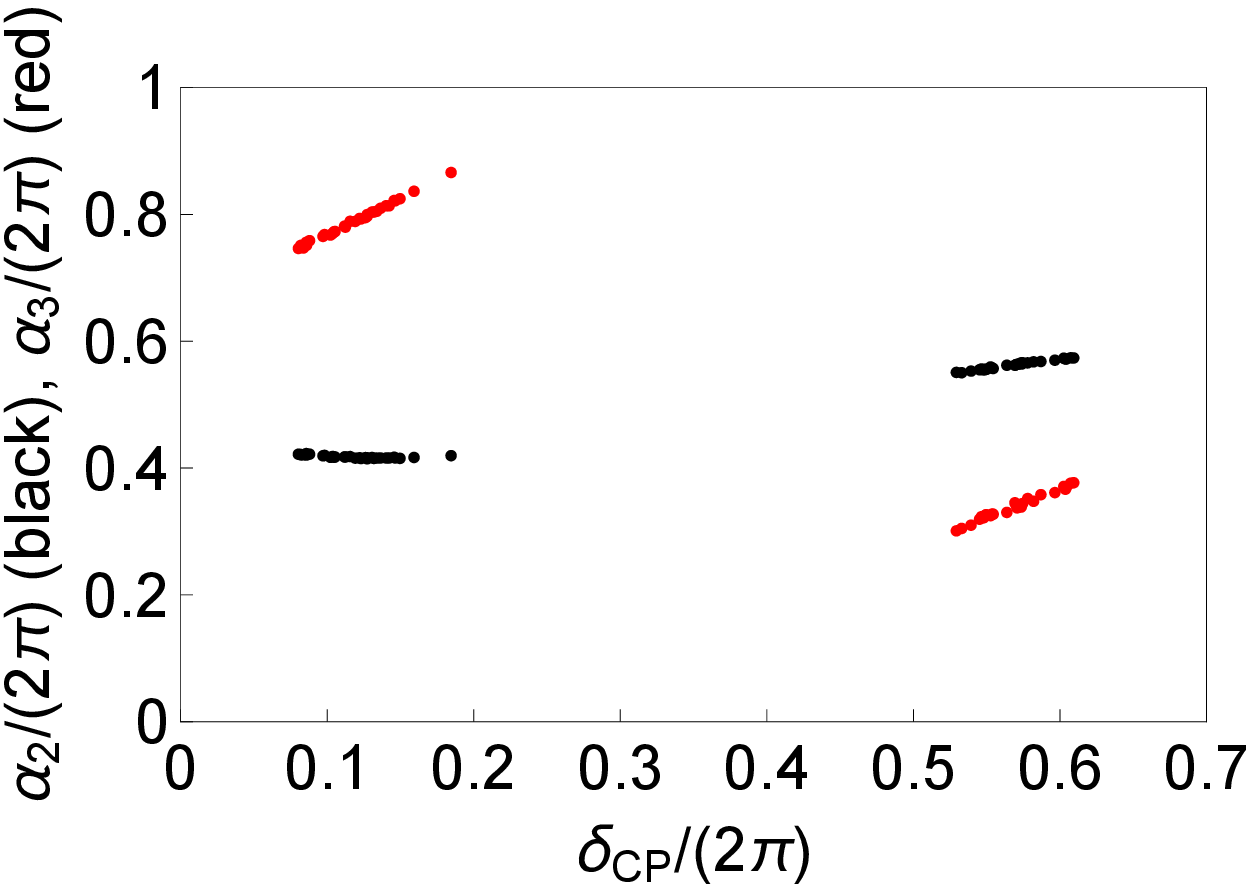}
\includegraphics[width=80mm]{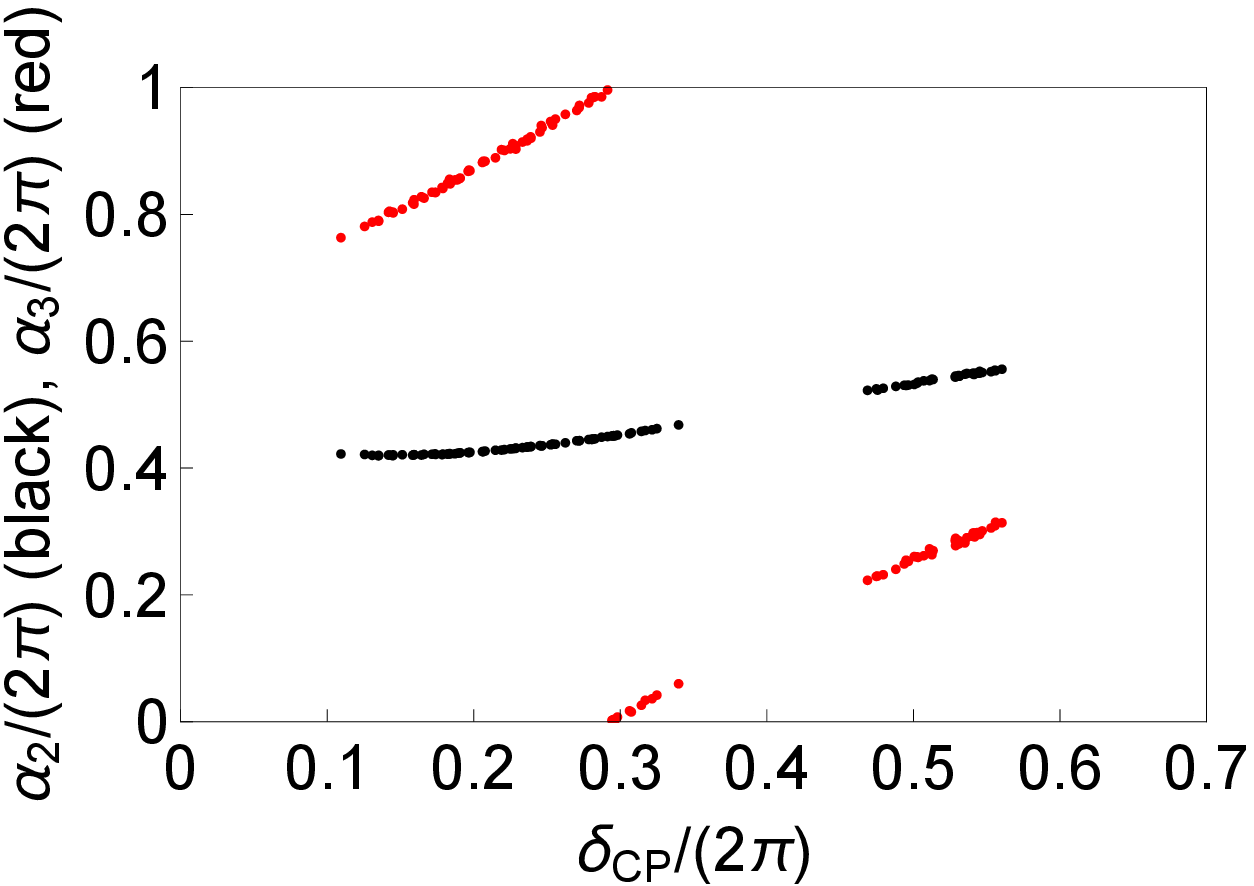}\\
\includegraphics[width=80mm]{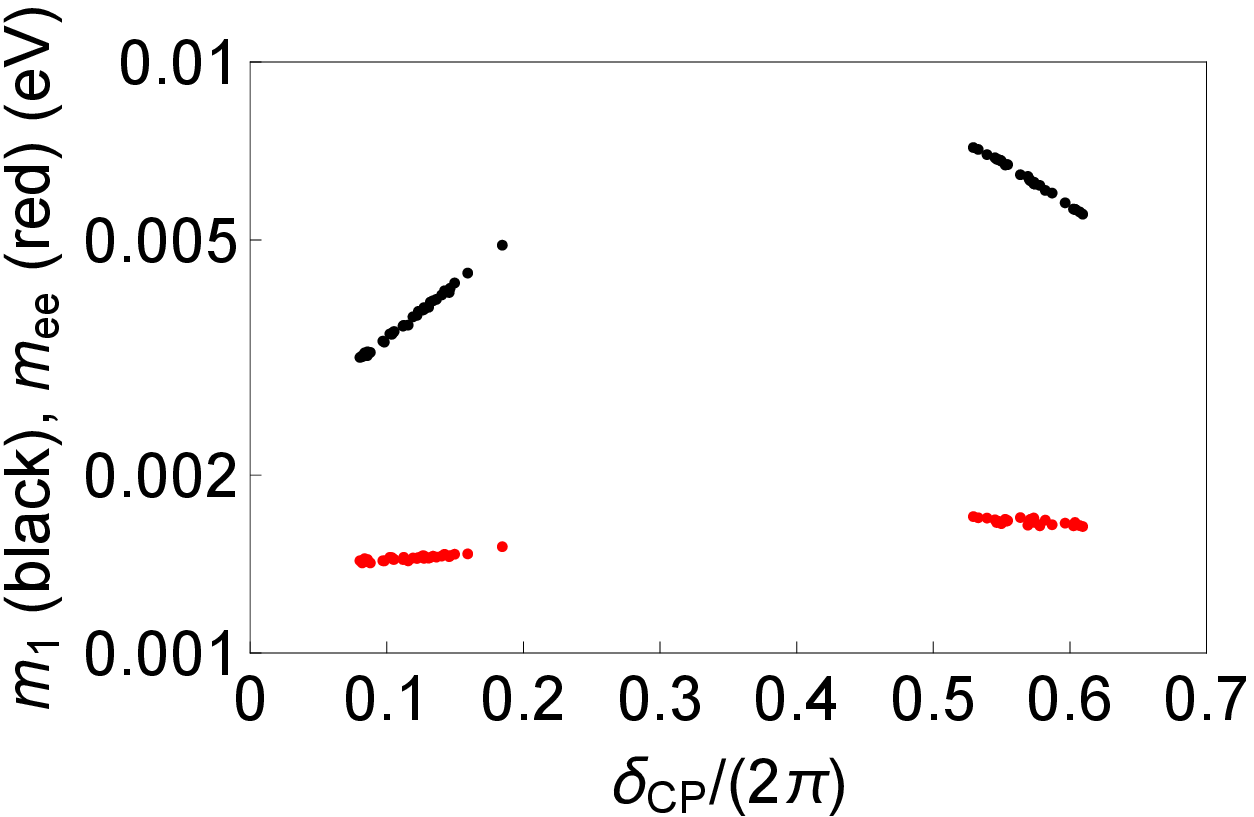}
\includegraphics[width=80mm]{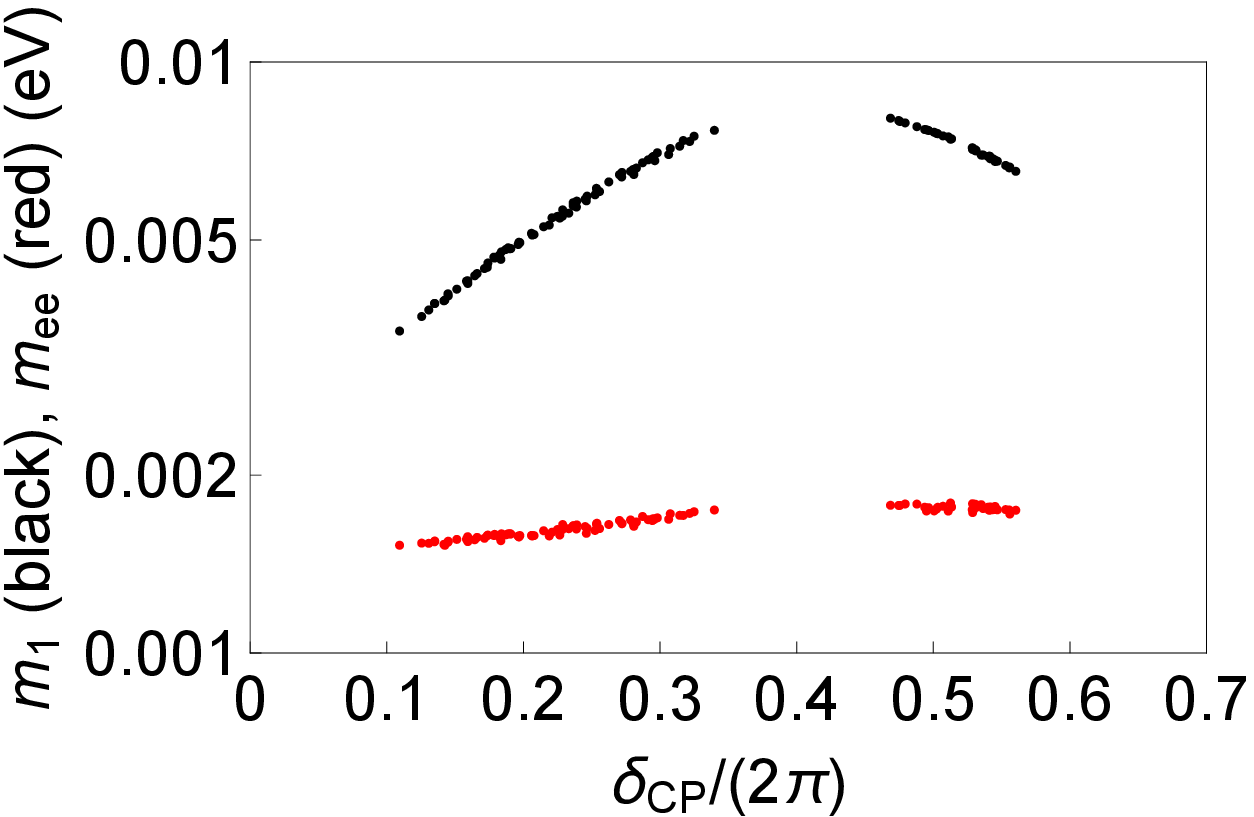}
\caption{Each dot represents a solution to Eq.~(\ref{pmnsckm}) that fits within the 2$\sigma$ range of
 $\theta_{12}^{ckm},\theta_{13}^{ckm},\theta_{23}^{ckm},\delta_{km}$
 at scale $\mu=10^{18}$~GeV evaluated by taking $\mu_{EW}=M_Z$.
These solutions are obtained from $1.6\times10^{10}$ random sets of values of $(\delta_{CP},\,\alpha_2,\,\alpha_3,\,\log(m_1))$ in the range Eq.~(\ref{randomrange}).
 The input values of $y_d,y_s,y_b$ and $\theta_{12},\theta_{13},\theta_{23},\vert\Delta m_{32}^2\vert,\Delta m_{21}^2$ are as shown in Table~\ref{inputs}, with the left plots corresponding to $\sin^2\theta_{23}=0.43$
 and the right plots to $\sin^2\theta_{23}=0.47$.
The plots in the first row are on $(\theta_{13,test}^{ckm},\,\delta_{CP})$ plane,
 those in the second row with black dots are on $(\delta_{CP},\,\alpha_2)$ plane,
 those in the second row with red dots are on $(\delta_{CP},\,\alpha_3)$ plane,
 those in the third row with black dots are on $(\delta_{CP}/(2\pi),\,m_1)$ plane,
 and those in the third row with red dots are on $(\delta_{CP}/(2\pi),\,m_{ee})$ plane.
}
\label{043047}
\end{center}
\end{figure}
\begin{figure}[H]
\begin{center}
\includegraphics[width=80mm]{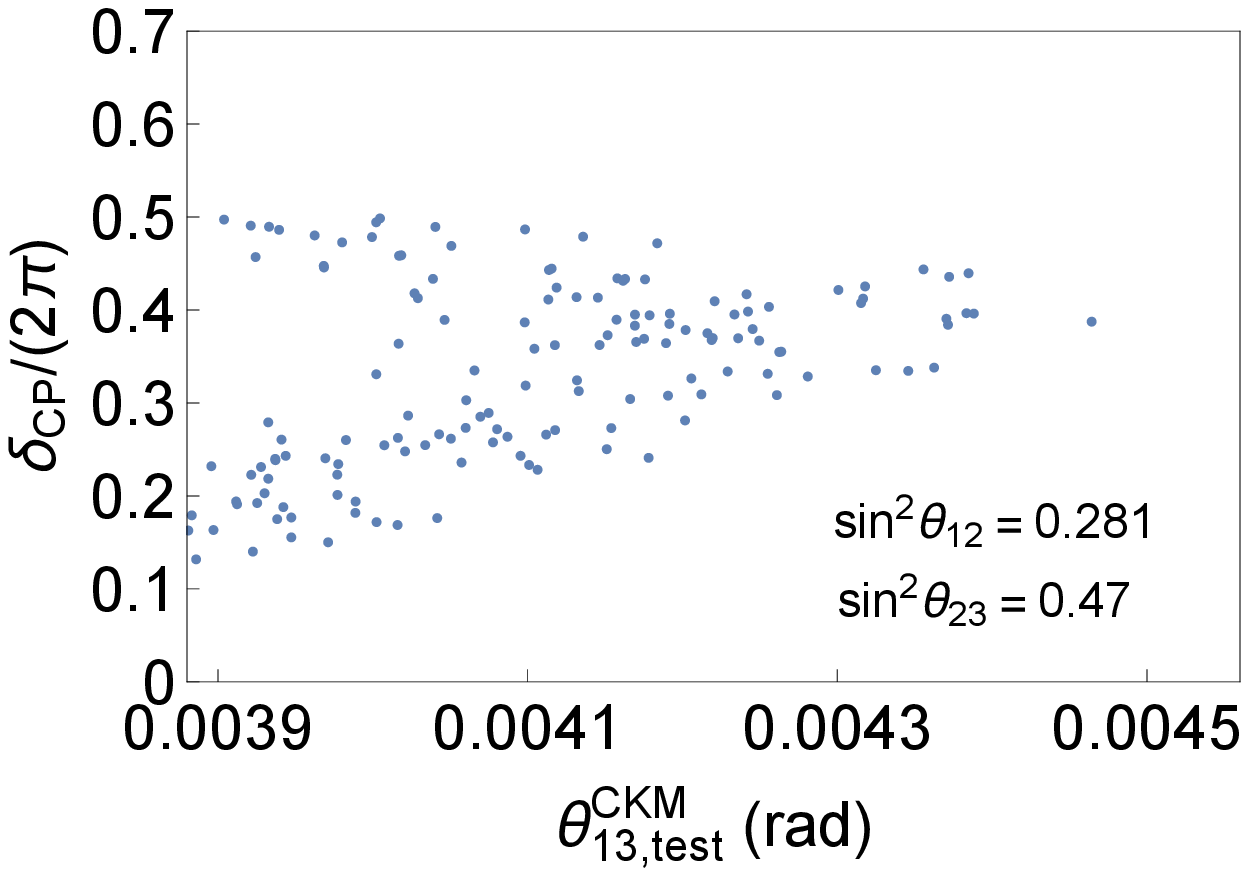}
\includegraphics[width=80mm]{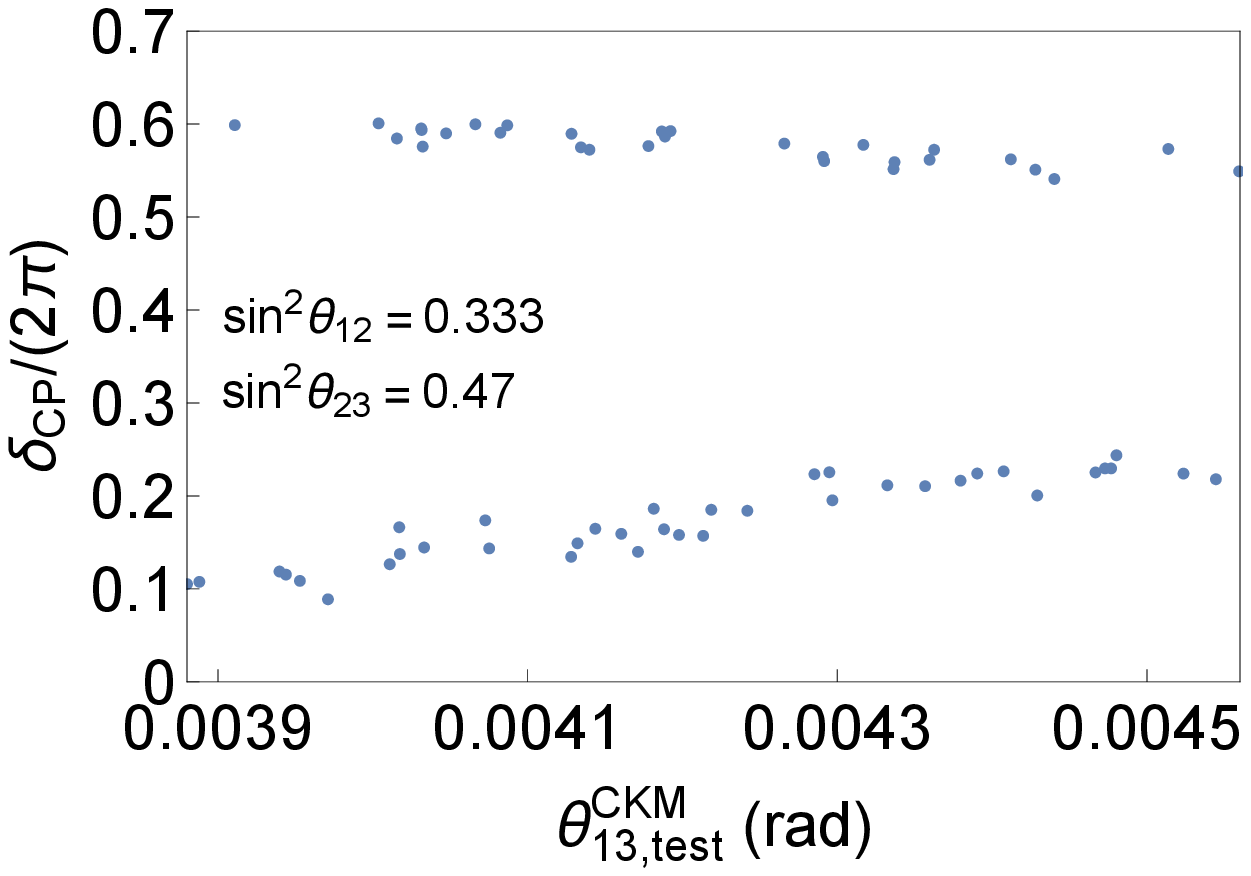}\\
\includegraphics[width=80mm]{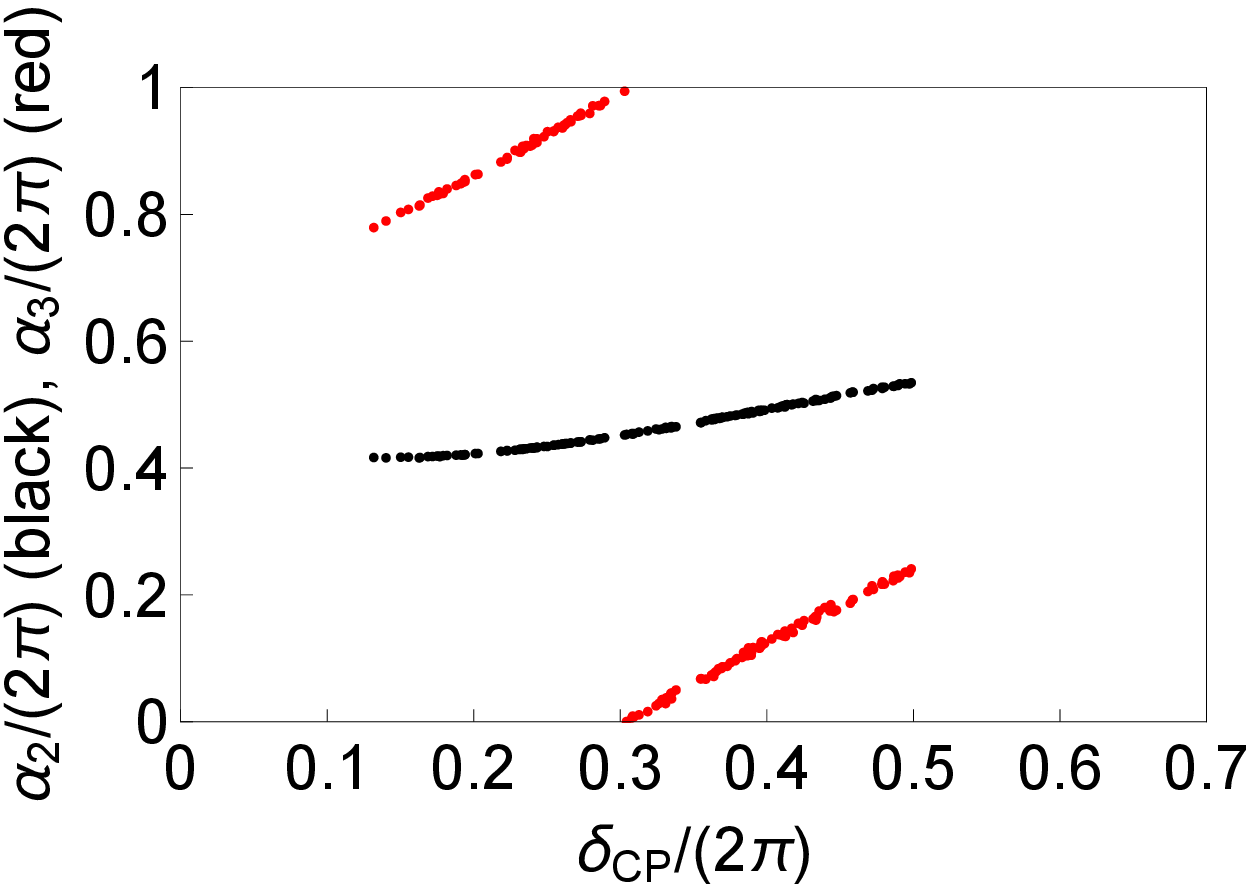}
\includegraphics[width=80mm]{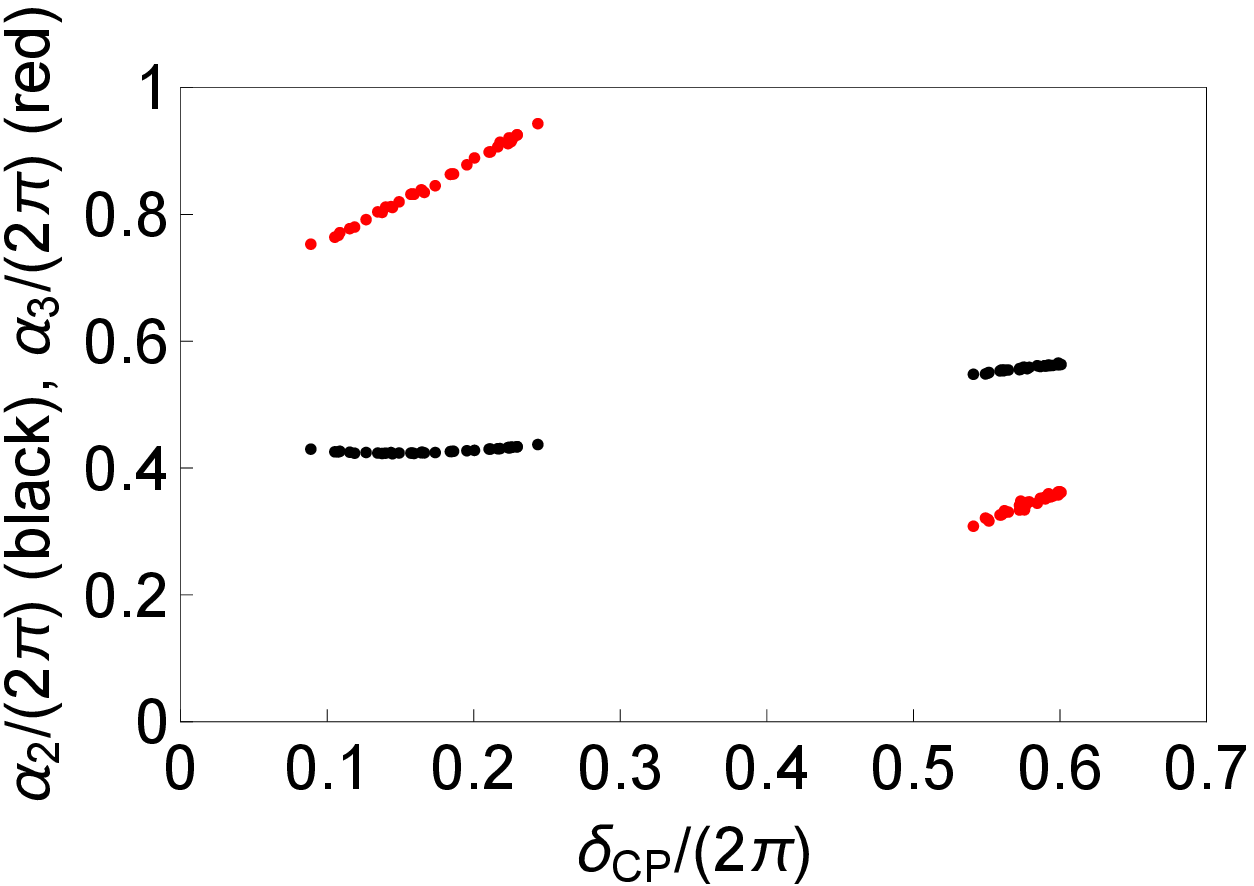}\\
\includegraphics[width=80mm]{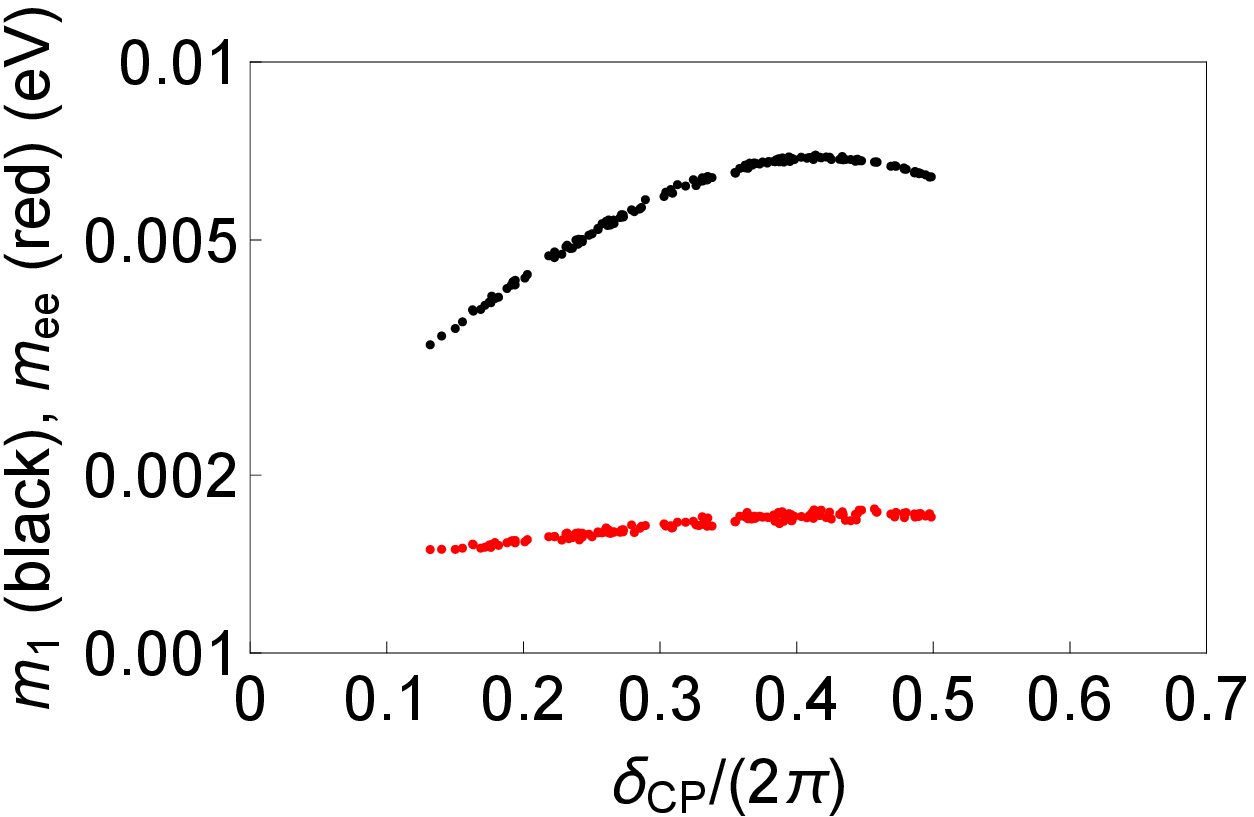}
\includegraphics[width=80mm]{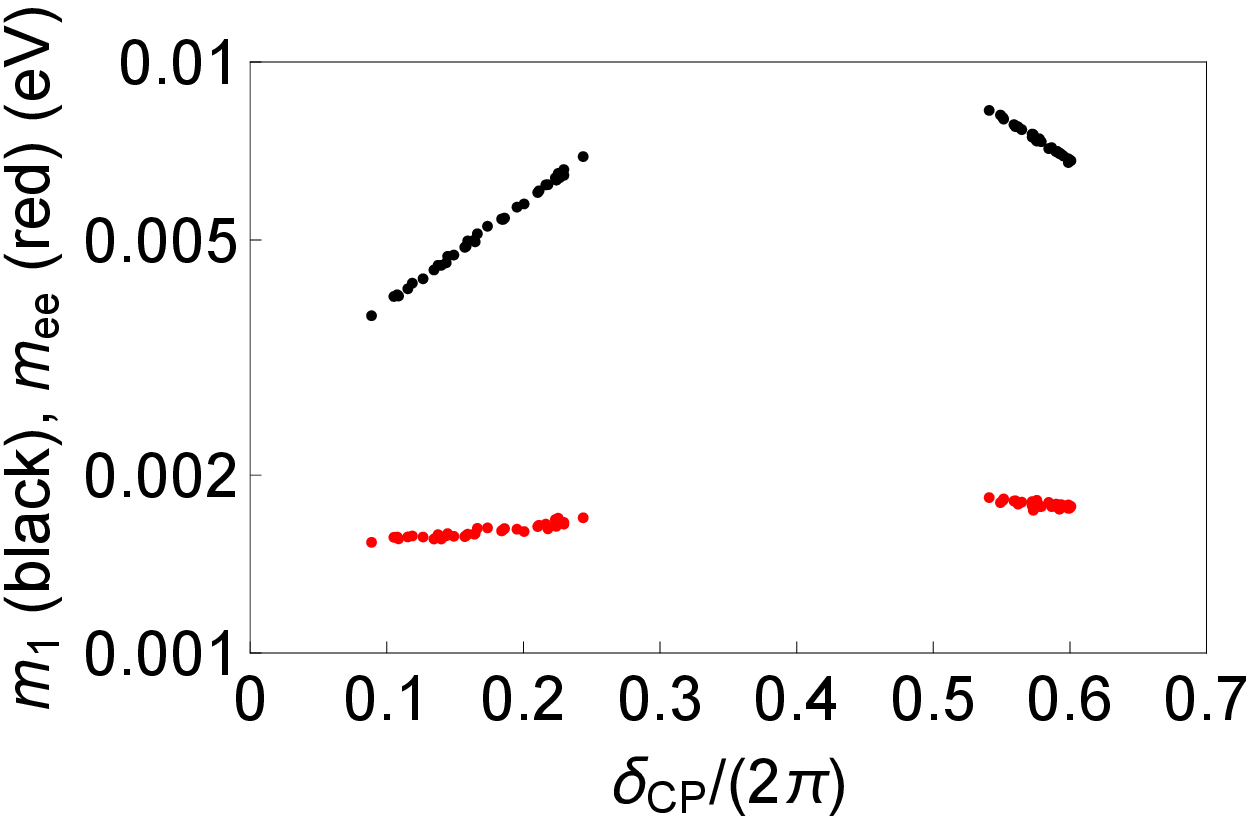}
\caption{Same as Figure~\ref{043047}, except that the different input values as shown in Table~\ref{inputs} are used, 
 with the left plots corresponding to $\sin^2\theta_{12}=0.281$
 and the right plots to $\sin^2\theta_{12}=0.333$.
}
\label{s12pm}
\end{center}
\end{figure}
\begin{figure}[H]
\begin{center}
\includegraphics[width=80mm]{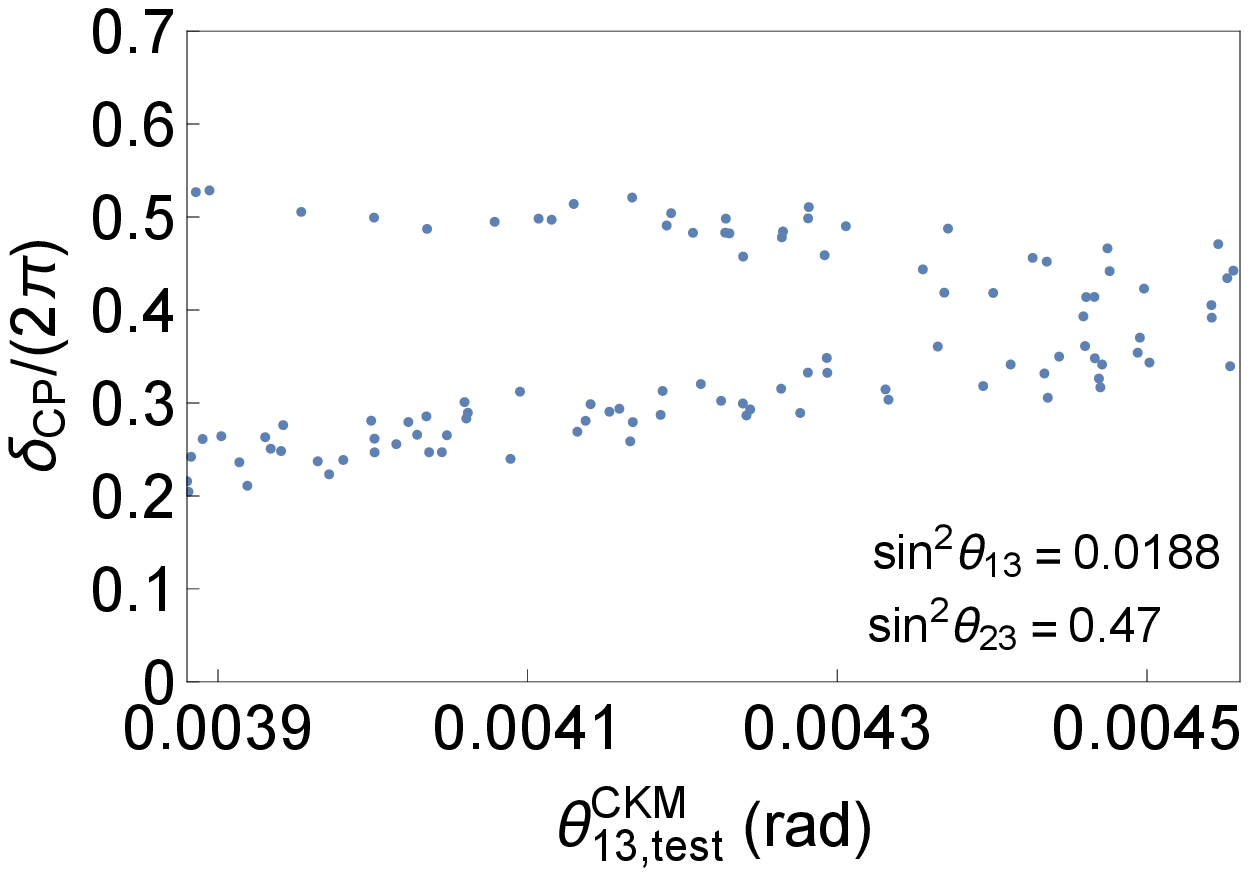}
\includegraphics[width=80mm]{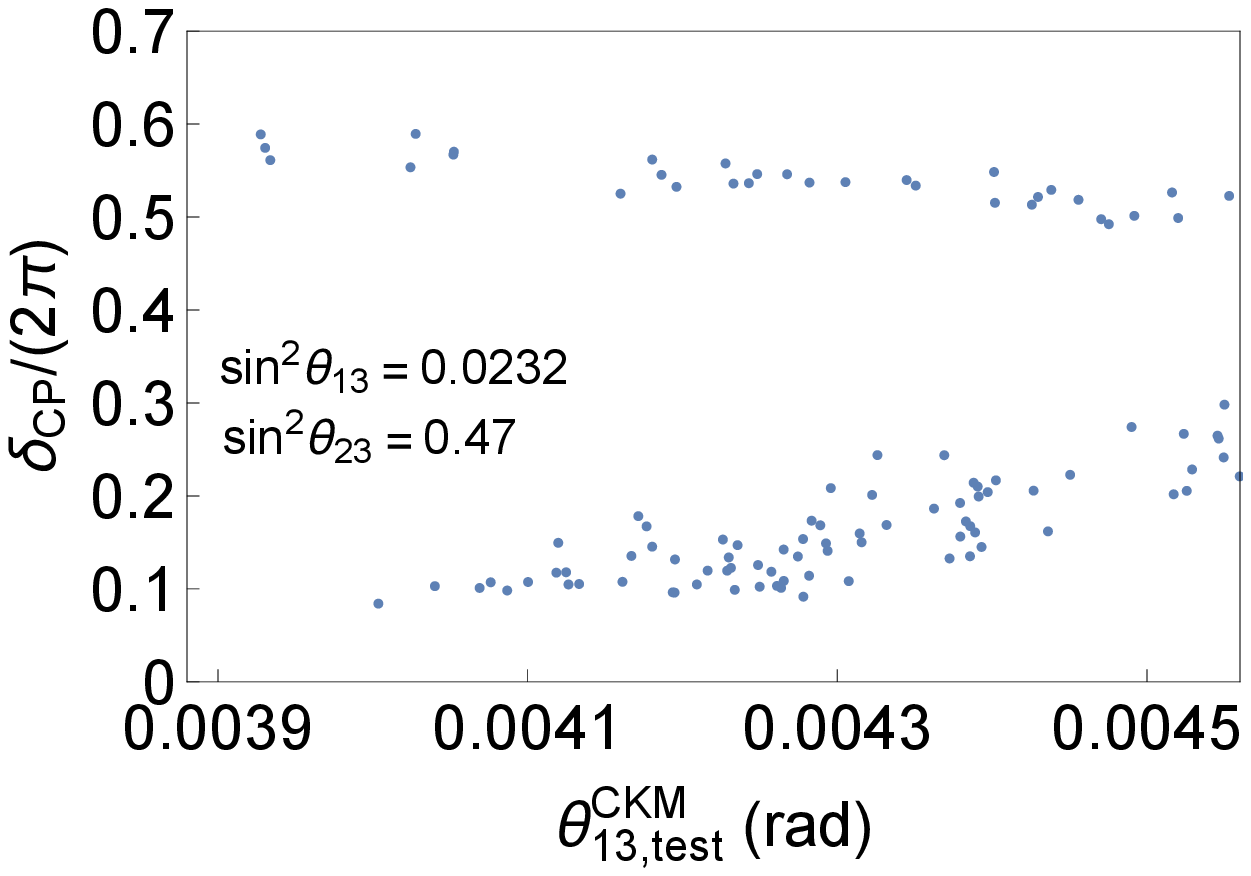}\\
\includegraphics[width=80mm]{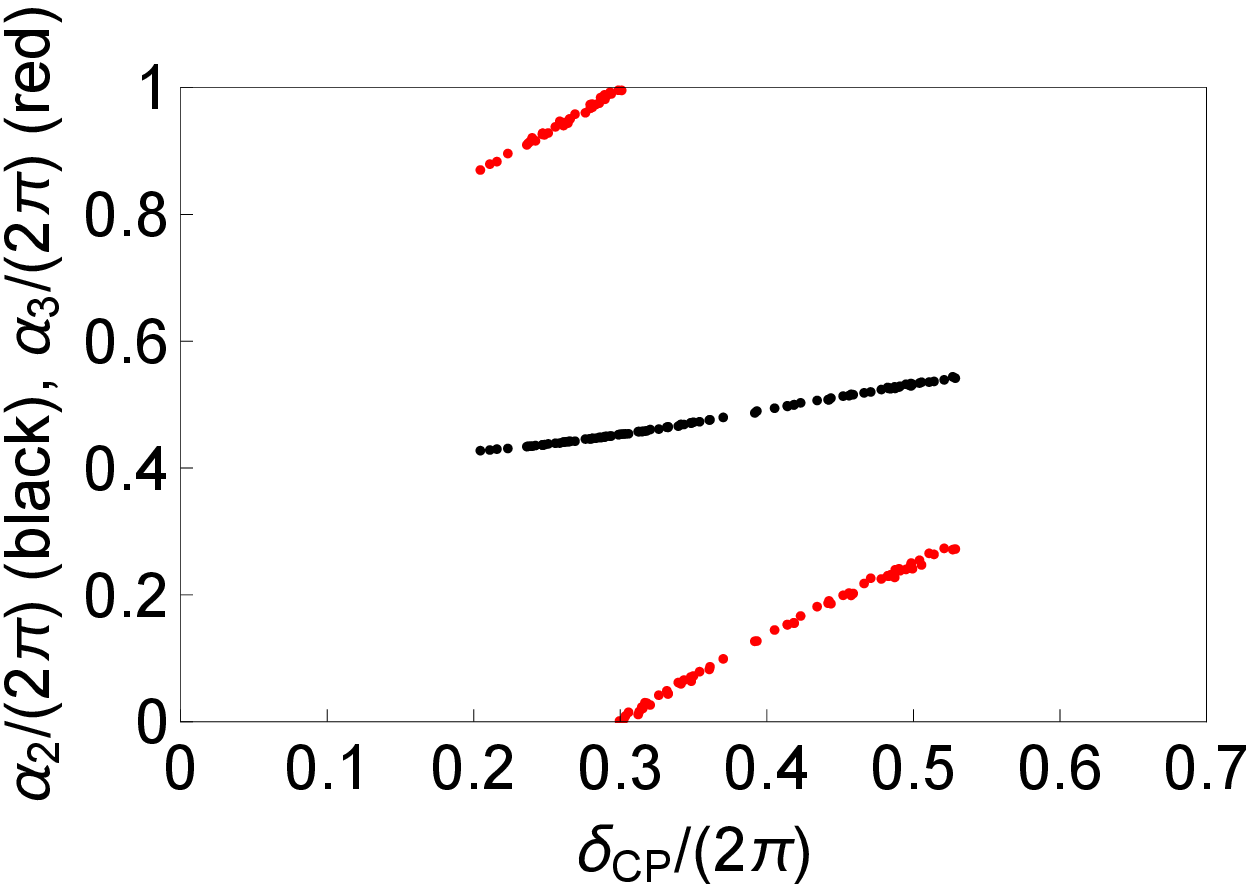}
\includegraphics[width=80mm]{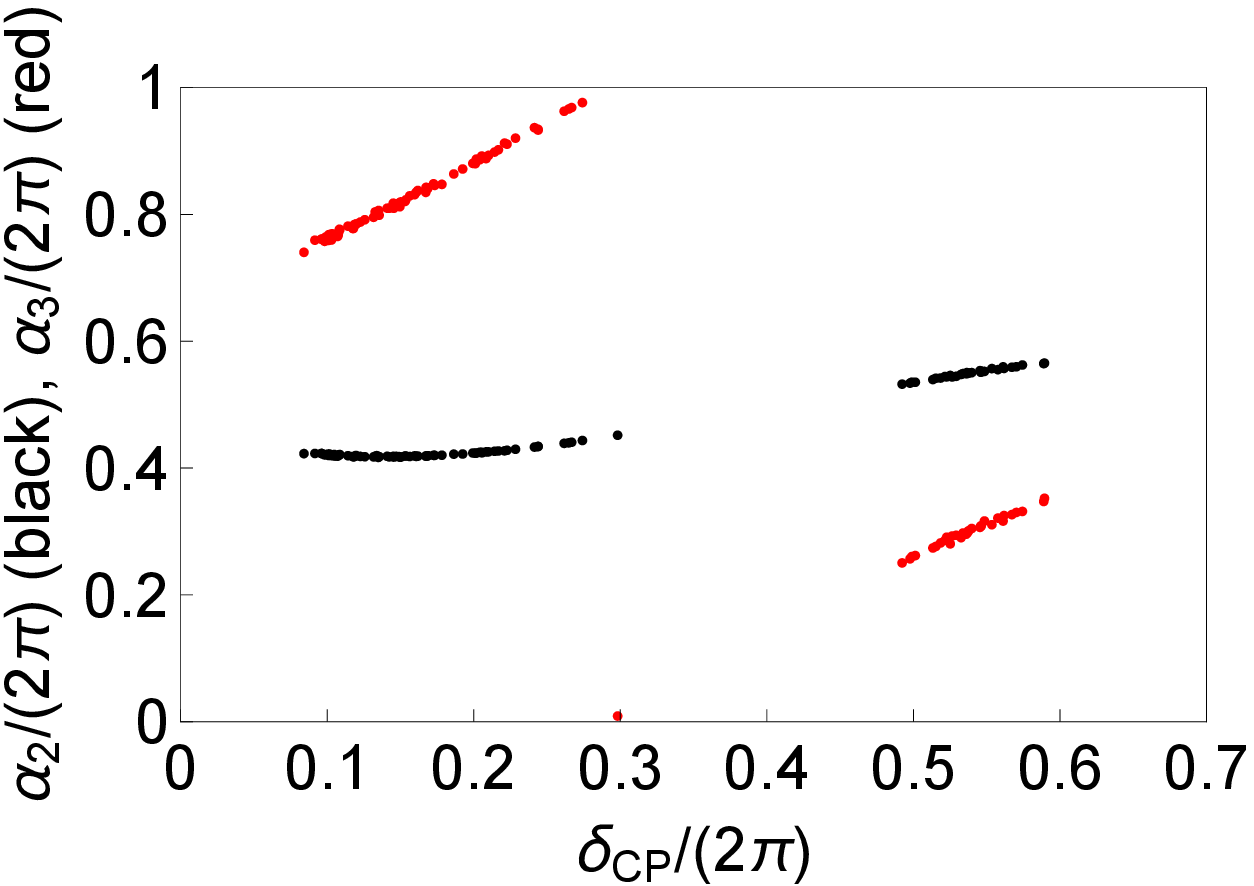}\\
\includegraphics[width=80mm]{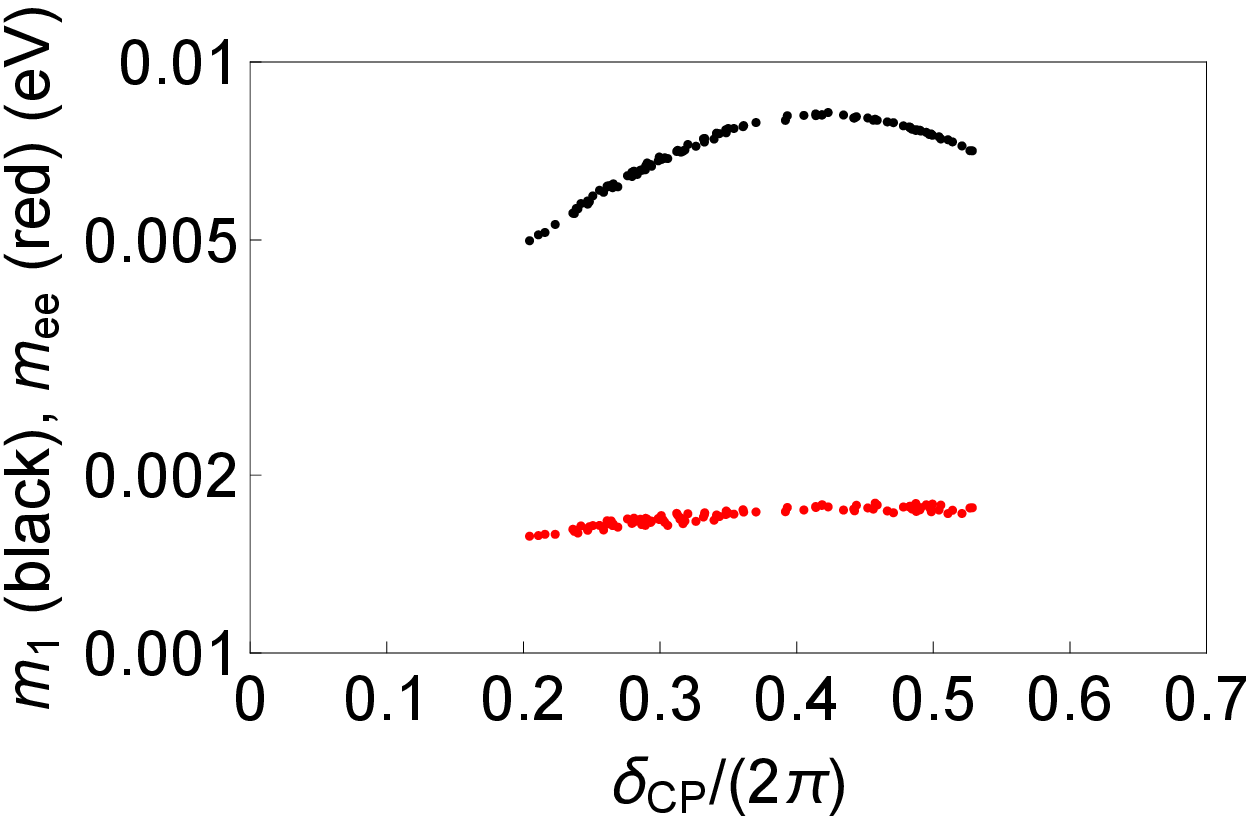}
\includegraphics[width=80mm]{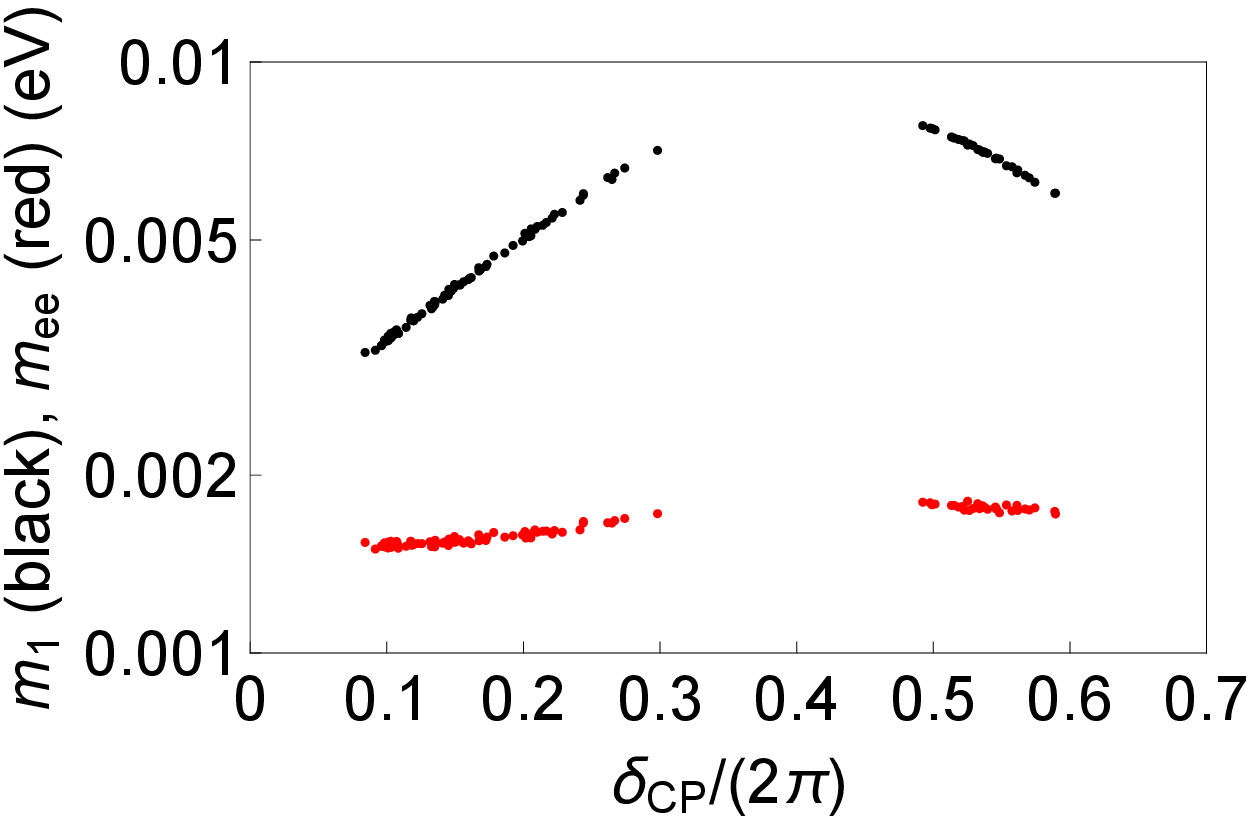}
\caption{Same as Figure~\ref{043047}, except that different input values as shown in Table~\ref{inputs} are used, 
 with the lefts plot corresponding to $\sin^2\theta_{13}=0.0188$
 and the right plots to $\sin^2\theta_{13}=0.0232$.
}
\label{s13pm}
\end{center}
\end{figure}
\begin{figure}[H]
\begin{center}
\includegraphics[width=80mm]{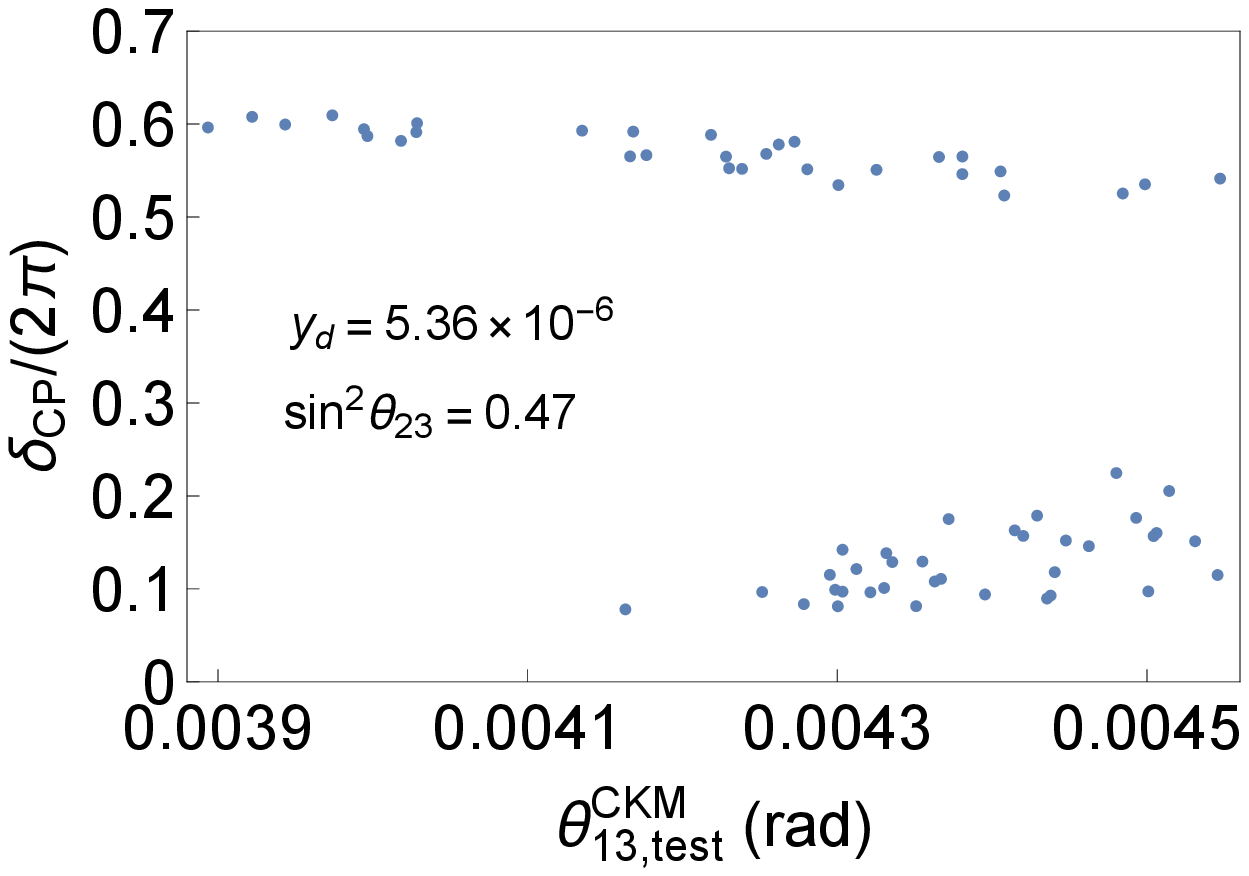}
\includegraphics[width=80mm]{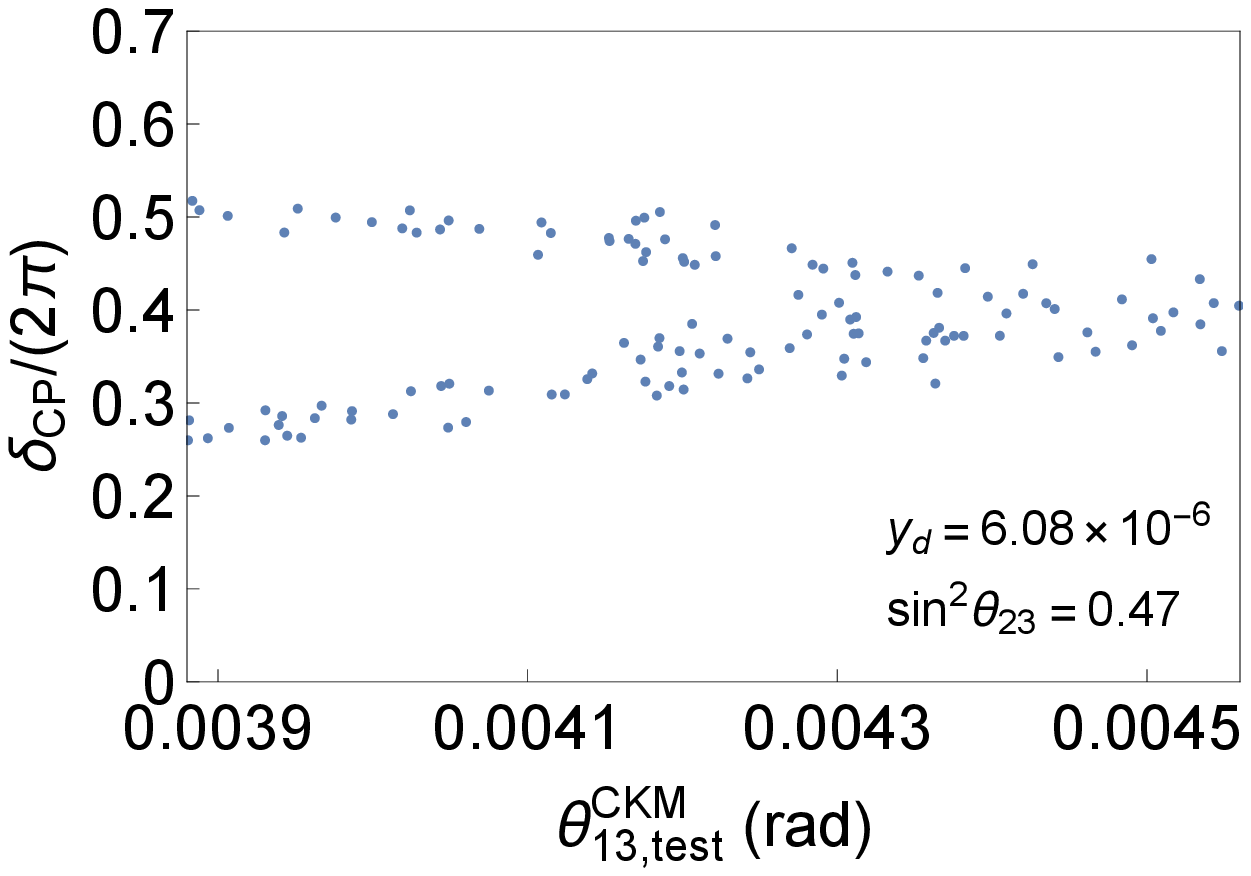}\\
\includegraphics[width=80mm]{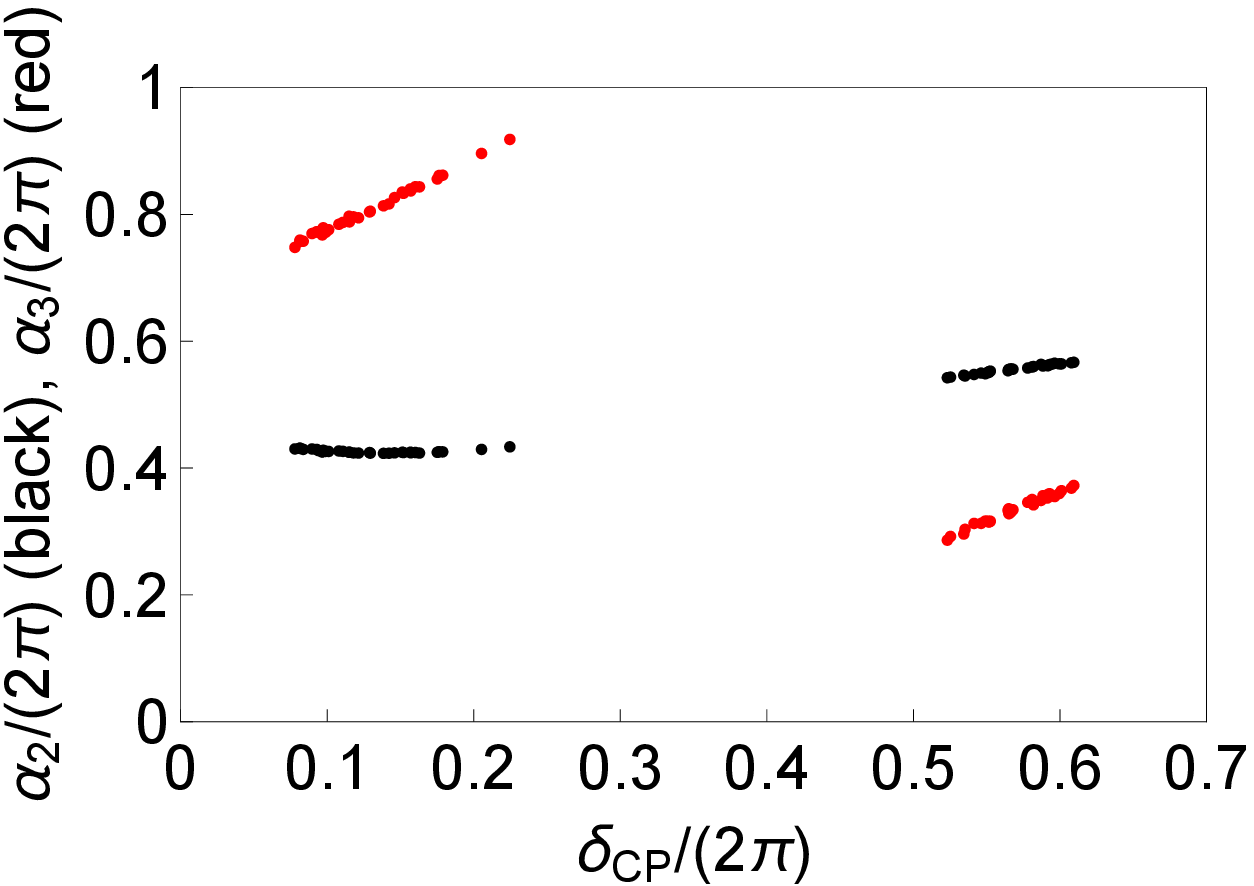}
\includegraphics[width=80mm]{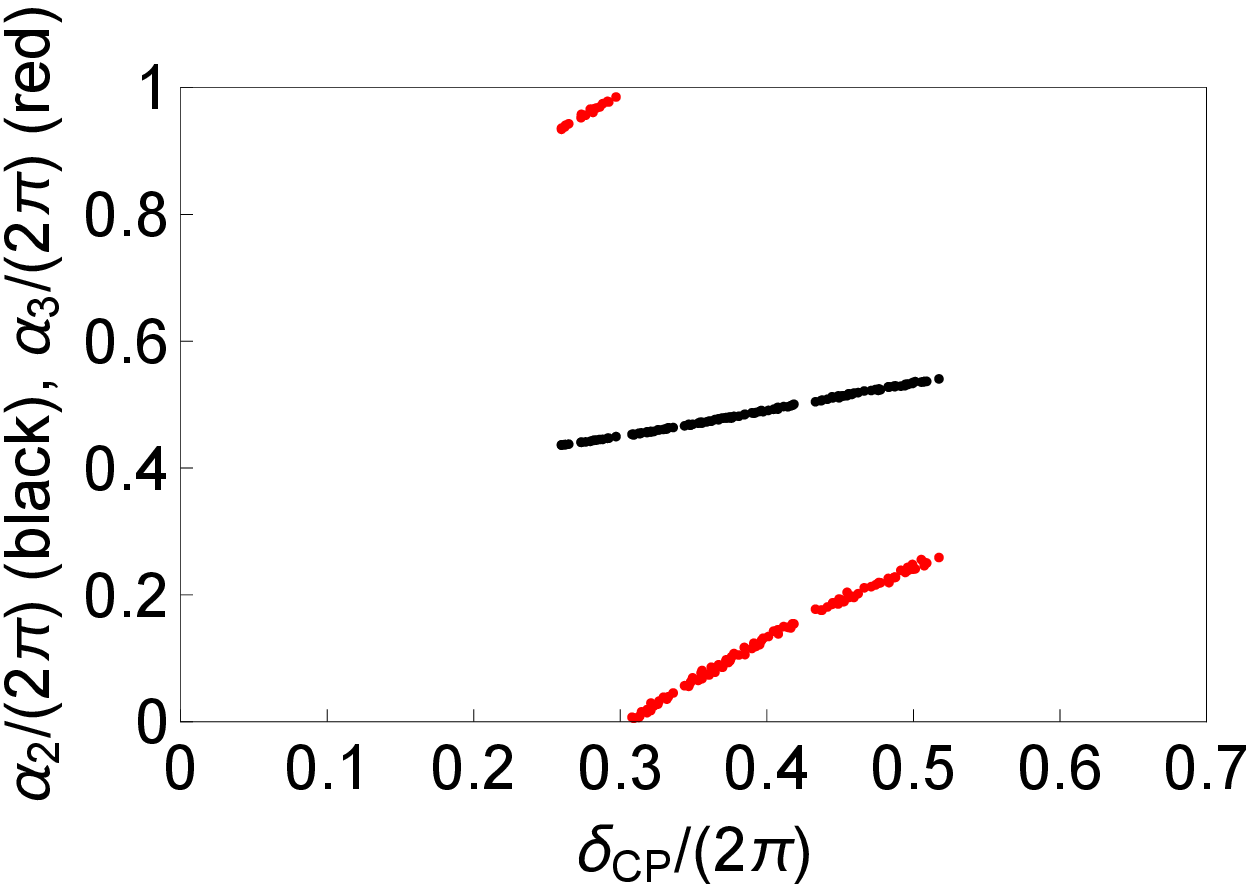}\\
\includegraphics[width=80mm]{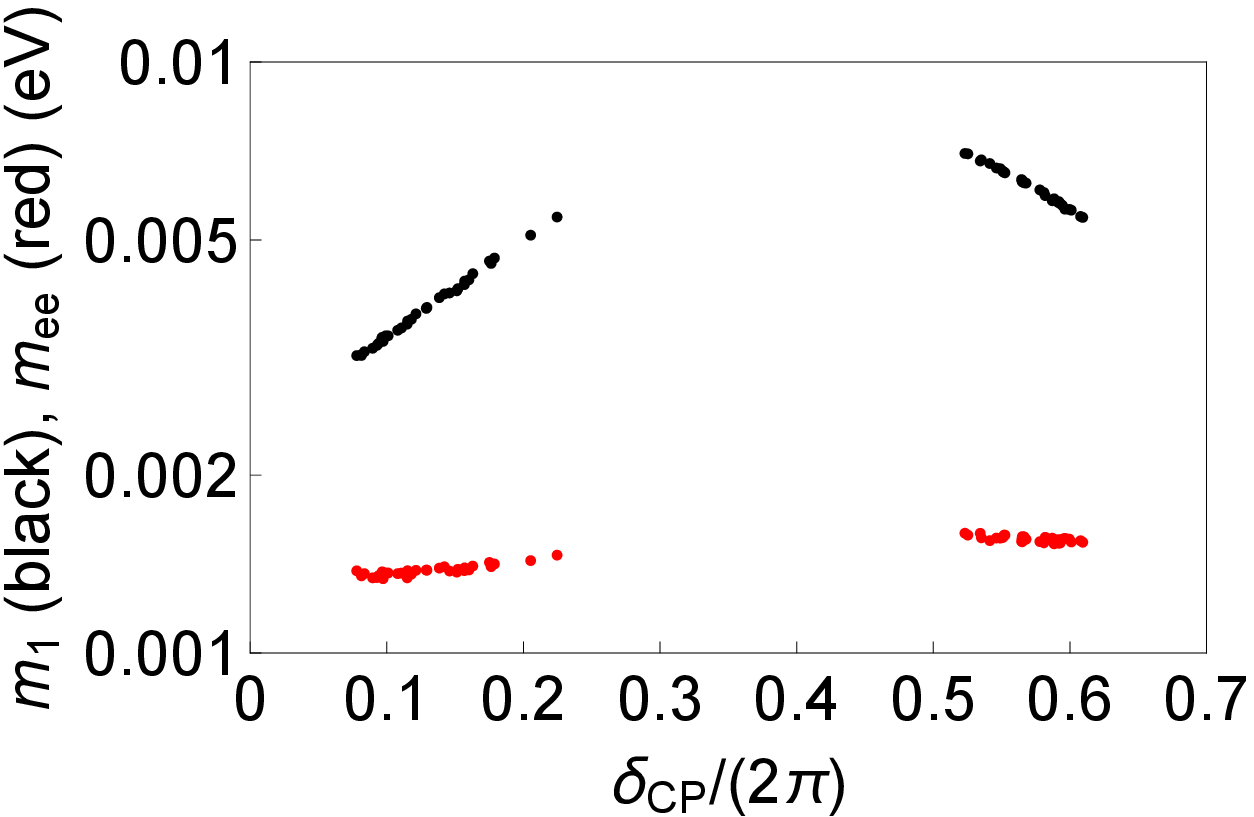}
\includegraphics[width=80mm]{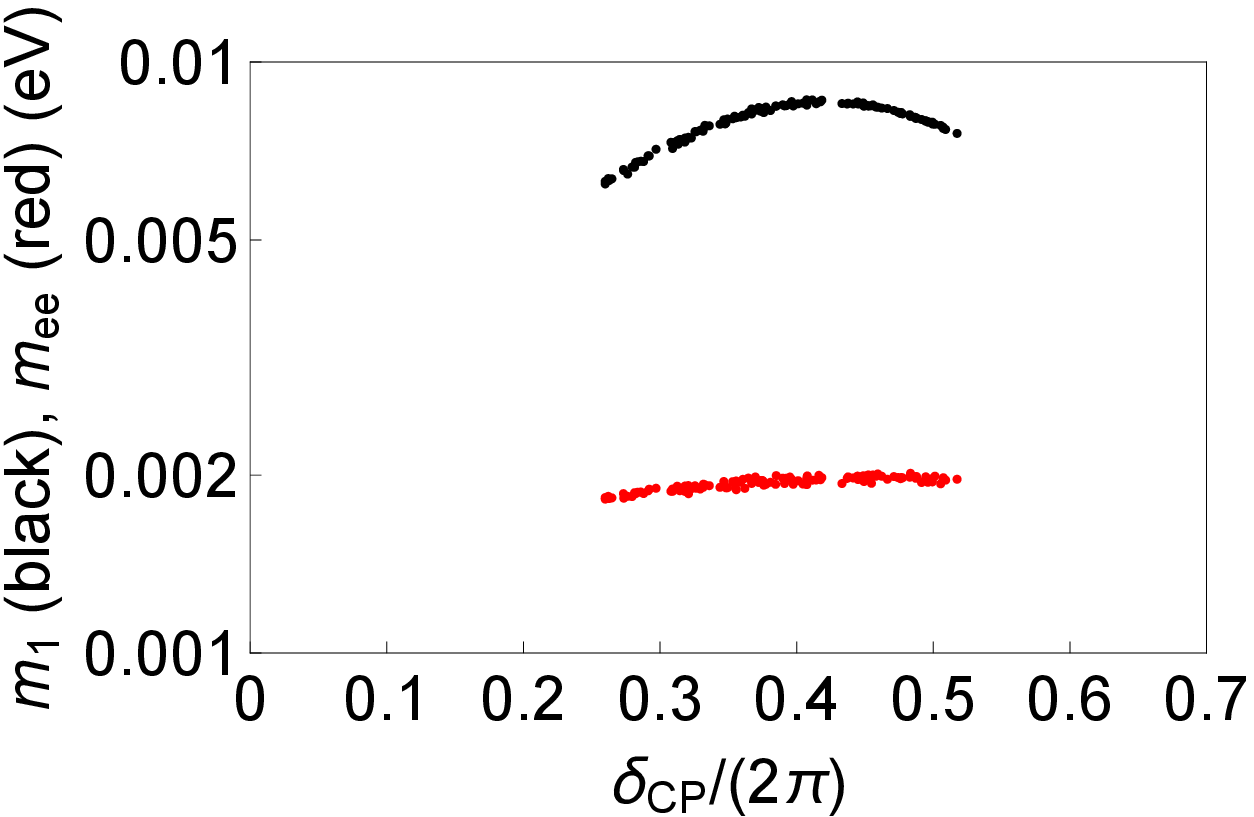}
\caption{Same as Figure~\ref{043047}, except that different input values as shown in Table~\ref{inputs} are used, 
 with the left plots corresponding to $y_d=5.36\times10^{-6}$
 and the right plots to $y_d=6.08\times10^{-6}$.
}
\label{ydpm}
\end{center}
\end{figure}
\begin{figure}[H]
\begin{center}
\includegraphics[width=80mm]{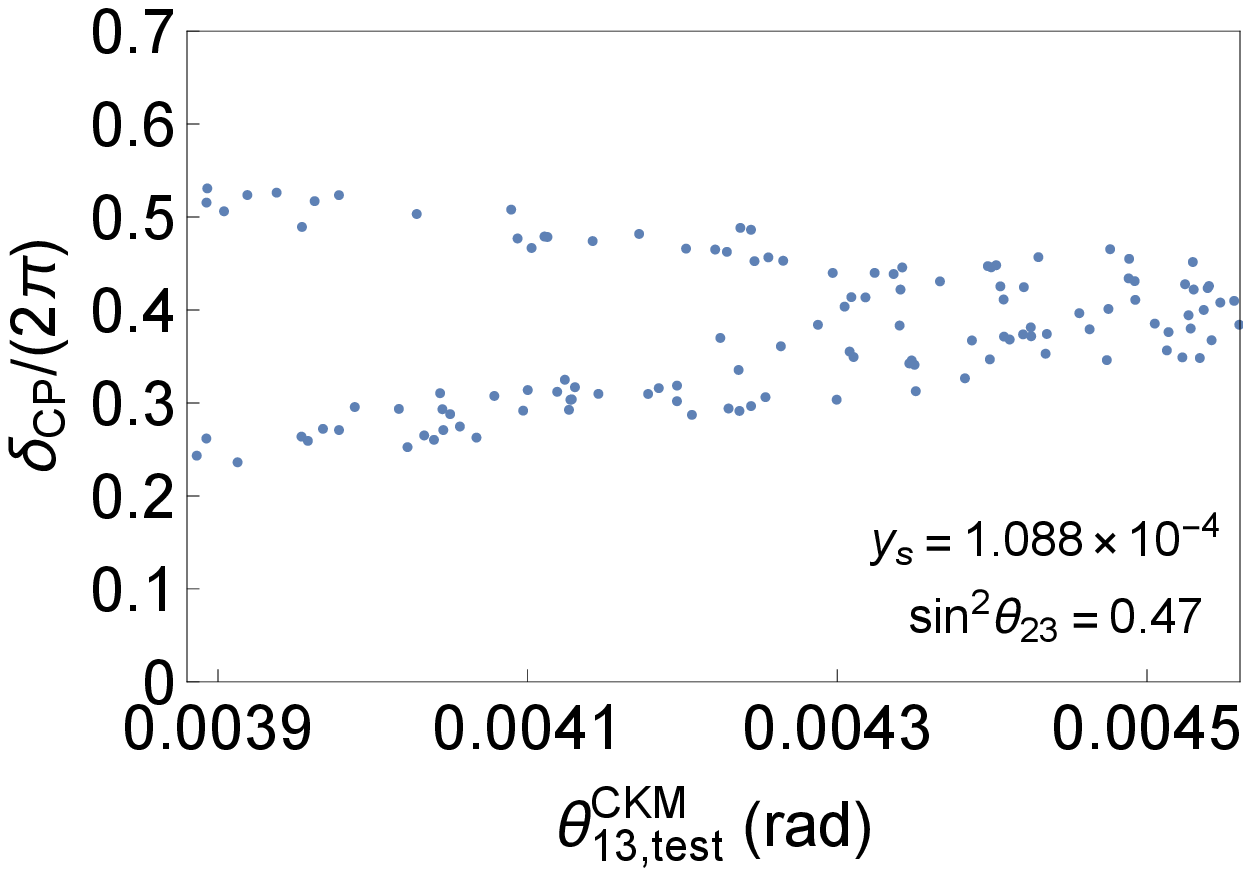}
\includegraphics[width=80mm]{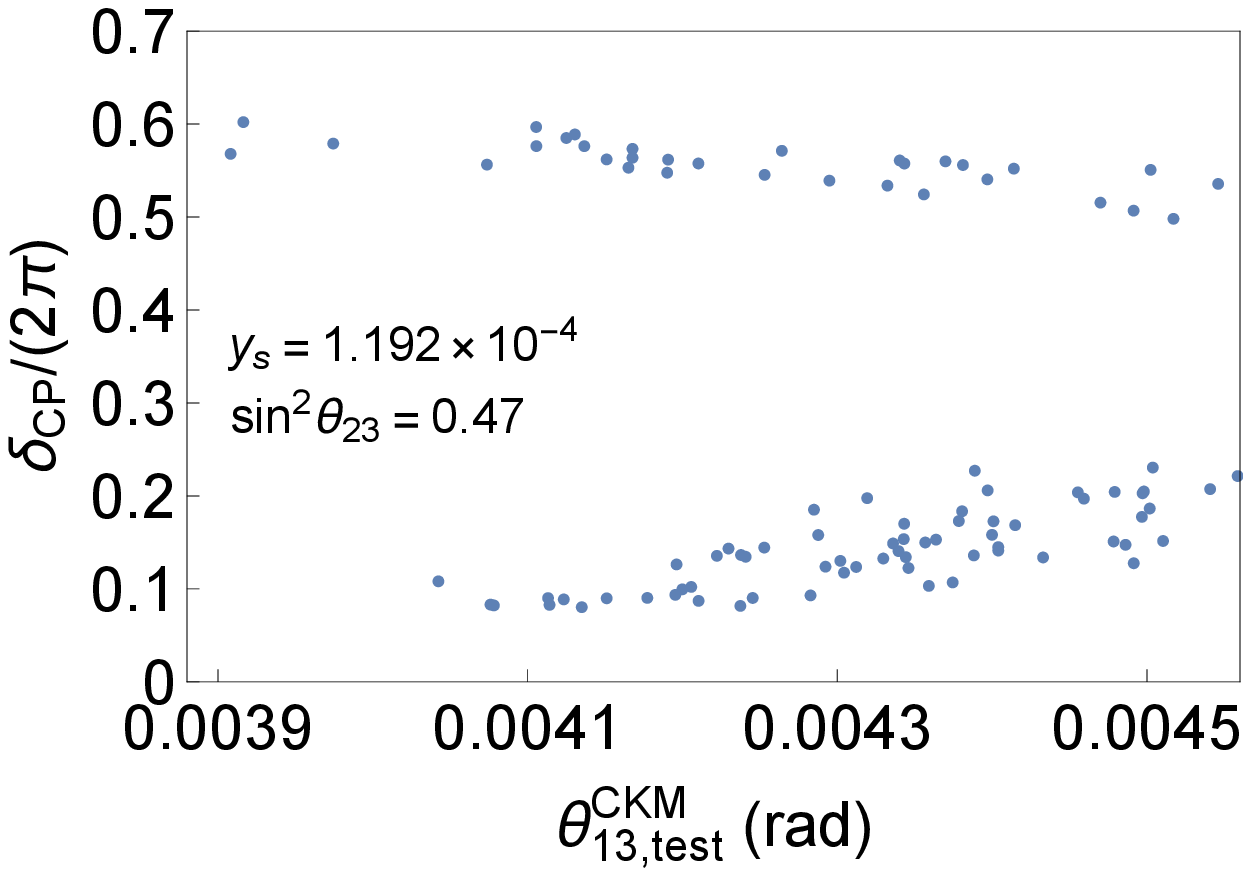}\\
\includegraphics[width=80mm]{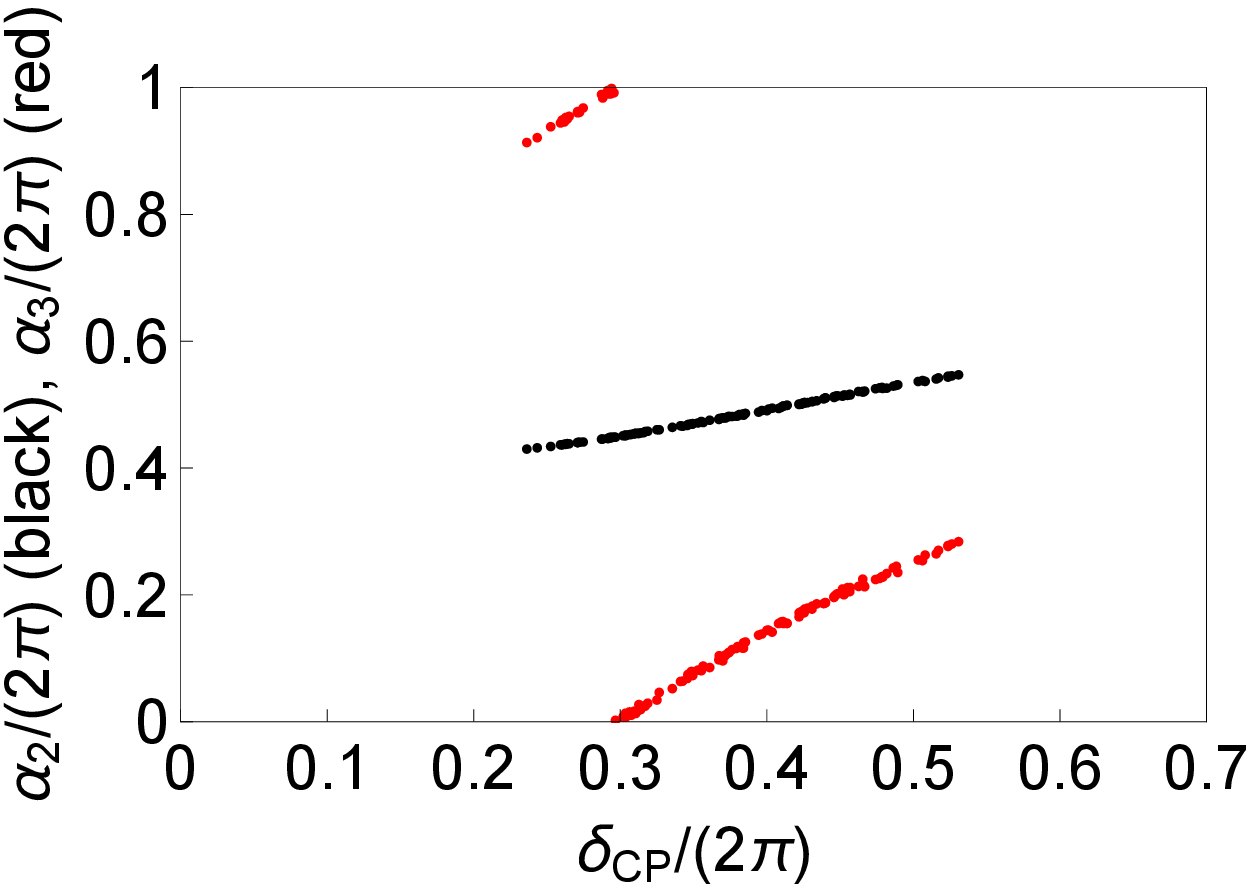}
\includegraphics[width=80mm]{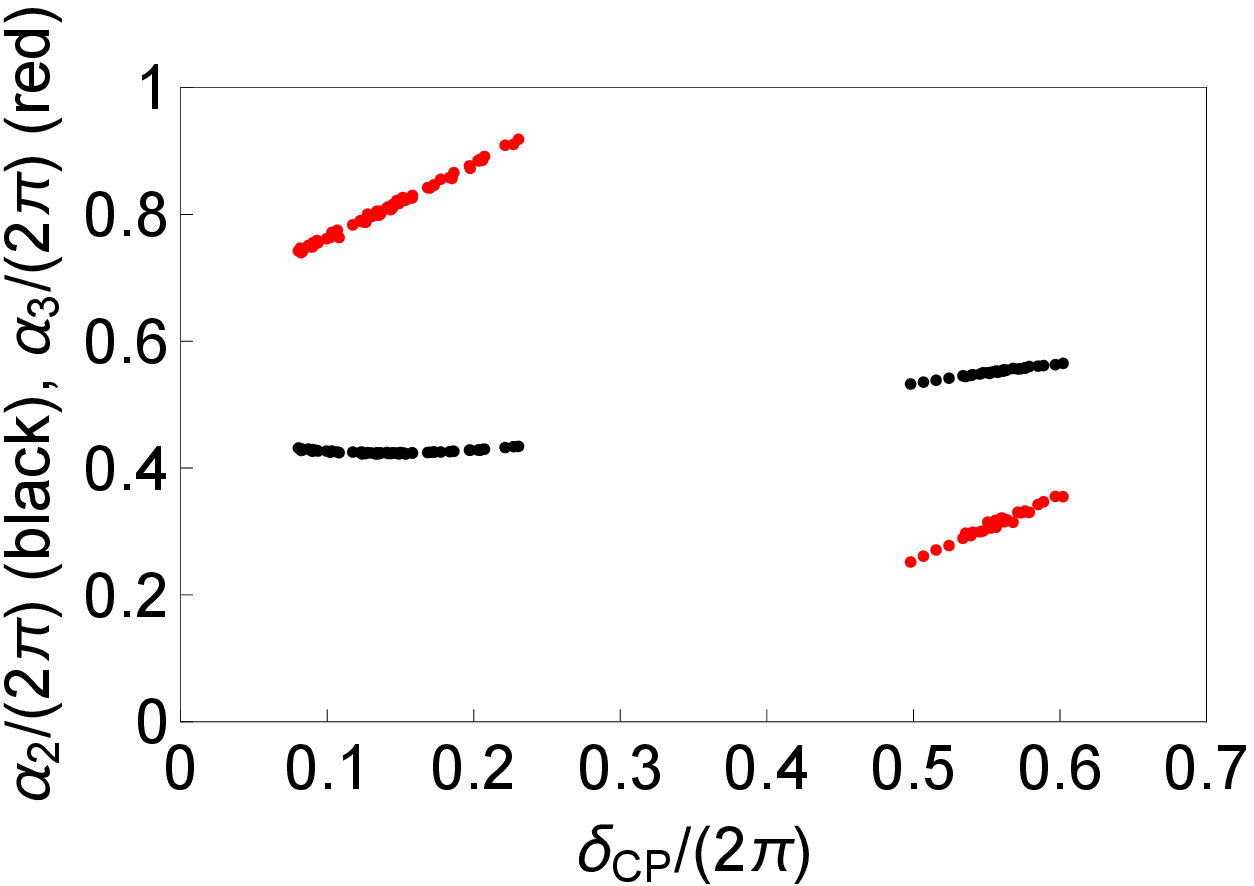}\\
\includegraphics[width=80mm]{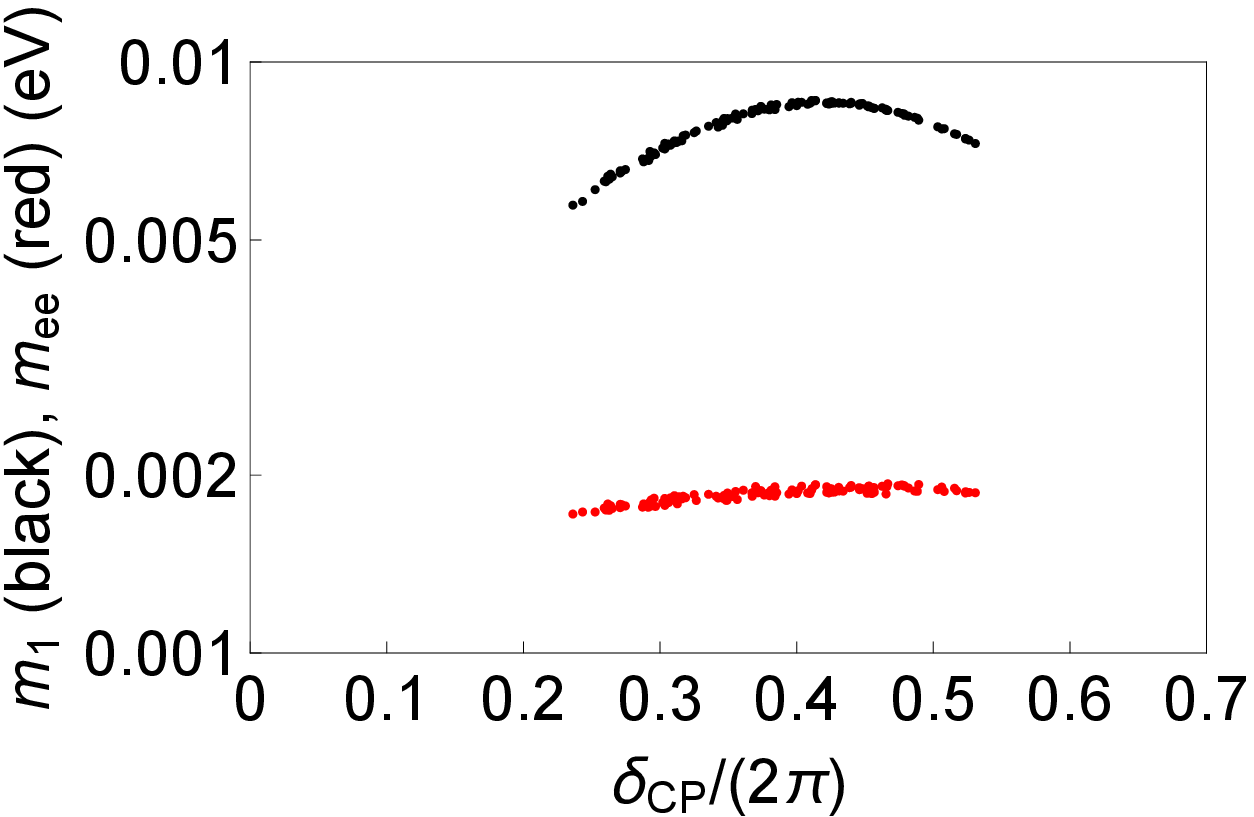}
\includegraphics[width=80mm]{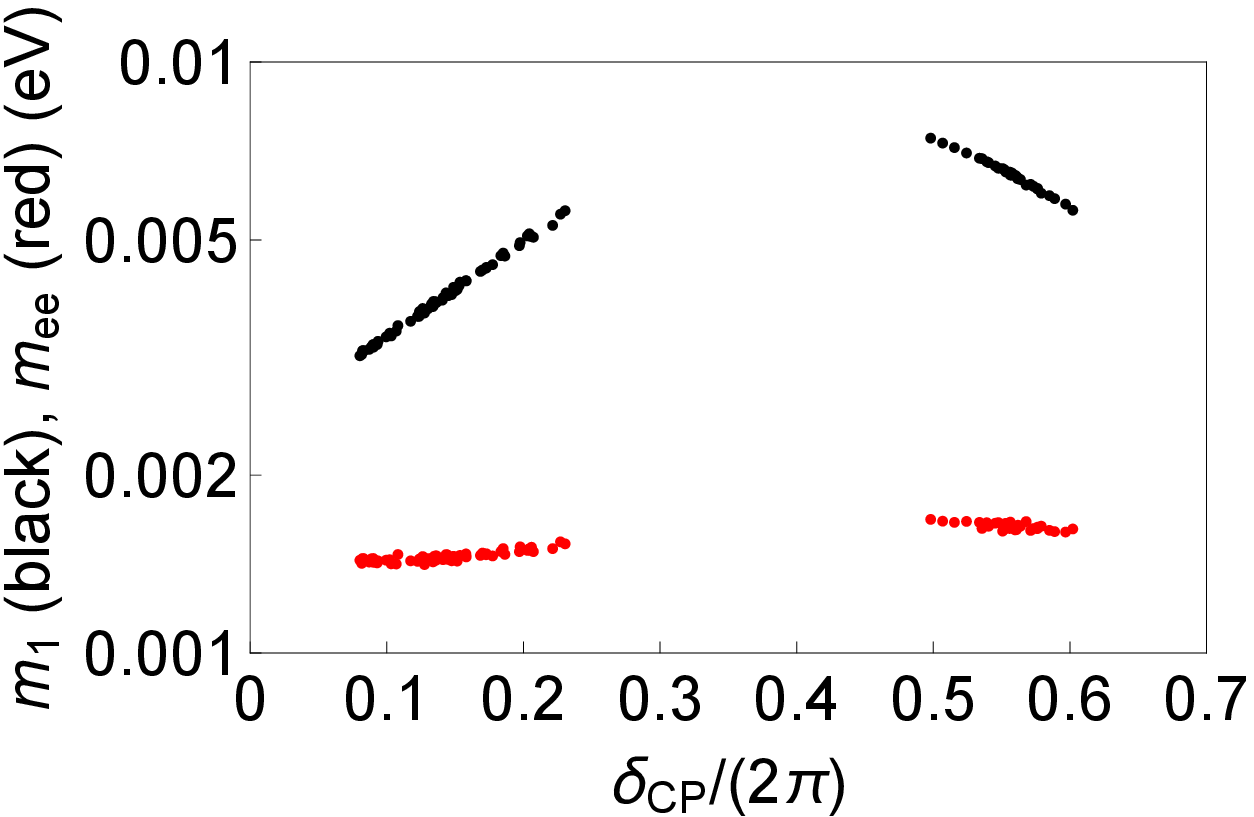}
\caption{Same as Figure~\ref{043047}, except that different input values as shown in Table~\ref{inputs} are used, 
 with the left plots corresponding to $y_s=1.088\times10^{-4}$
 and the right plots to $y_s=1.192\times10^{-4}$.
}
\label{yspm}
\end{center}
\end{figure}
\begin{figure}[H]
\begin{center}
\includegraphics[width=80mm]{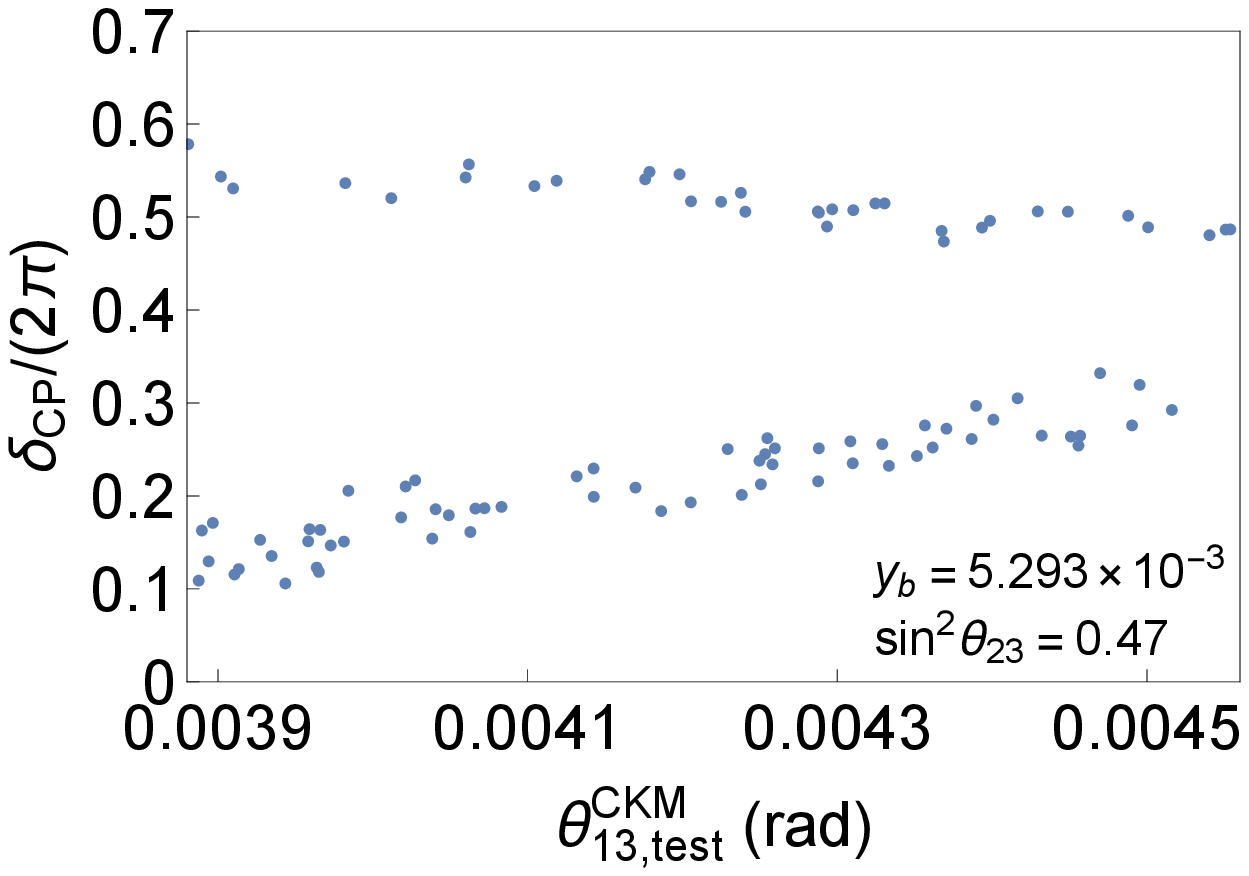}
\includegraphics[width=80mm]{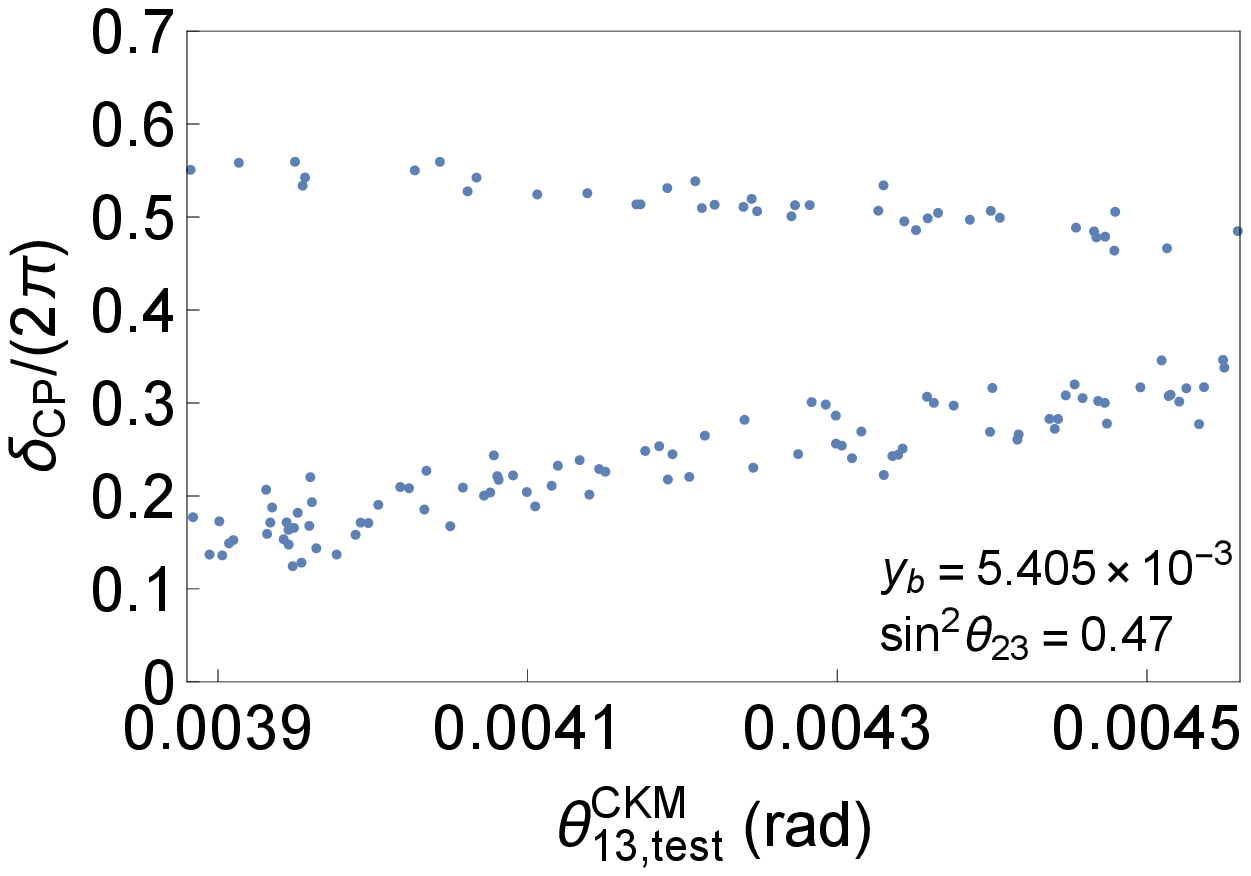}\\
\includegraphics[width=80mm]{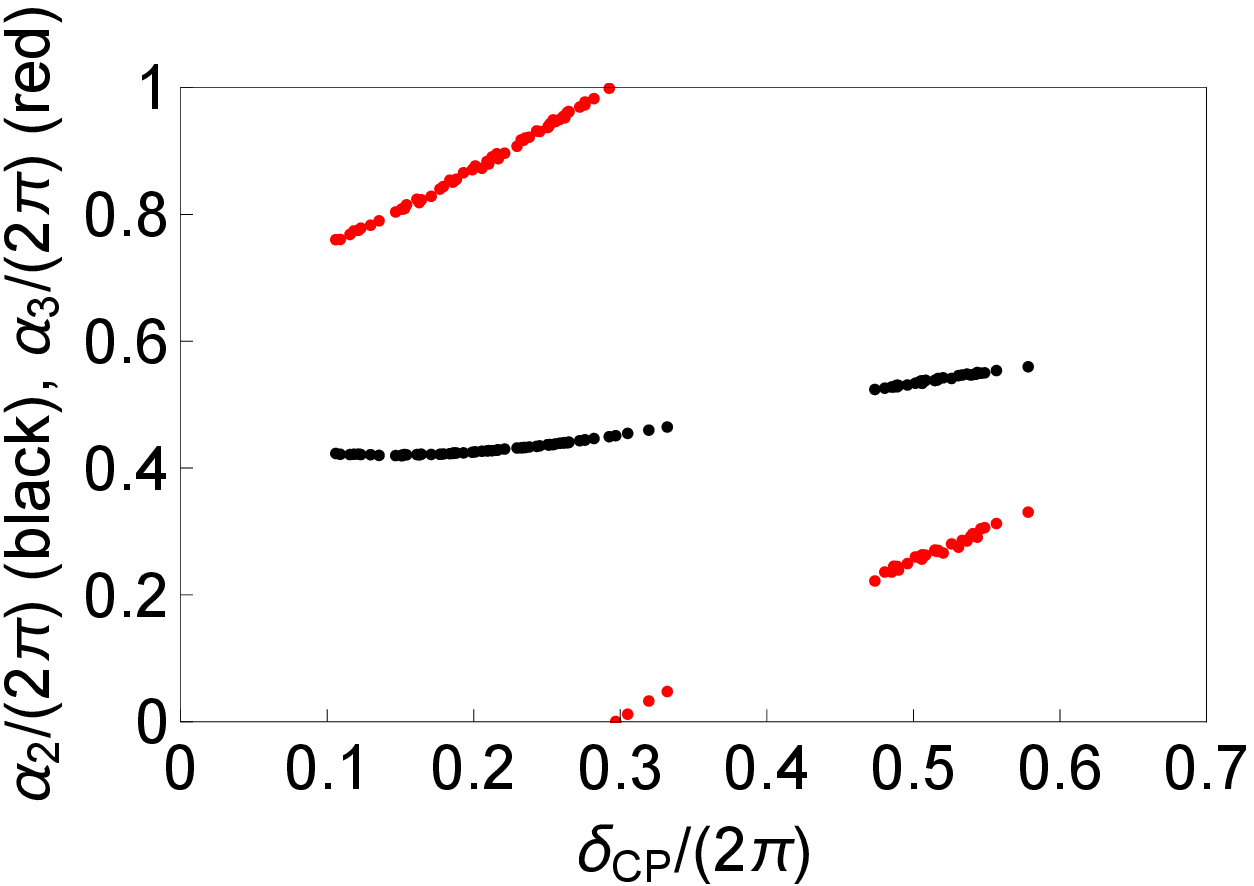}
\includegraphics[width=80mm]{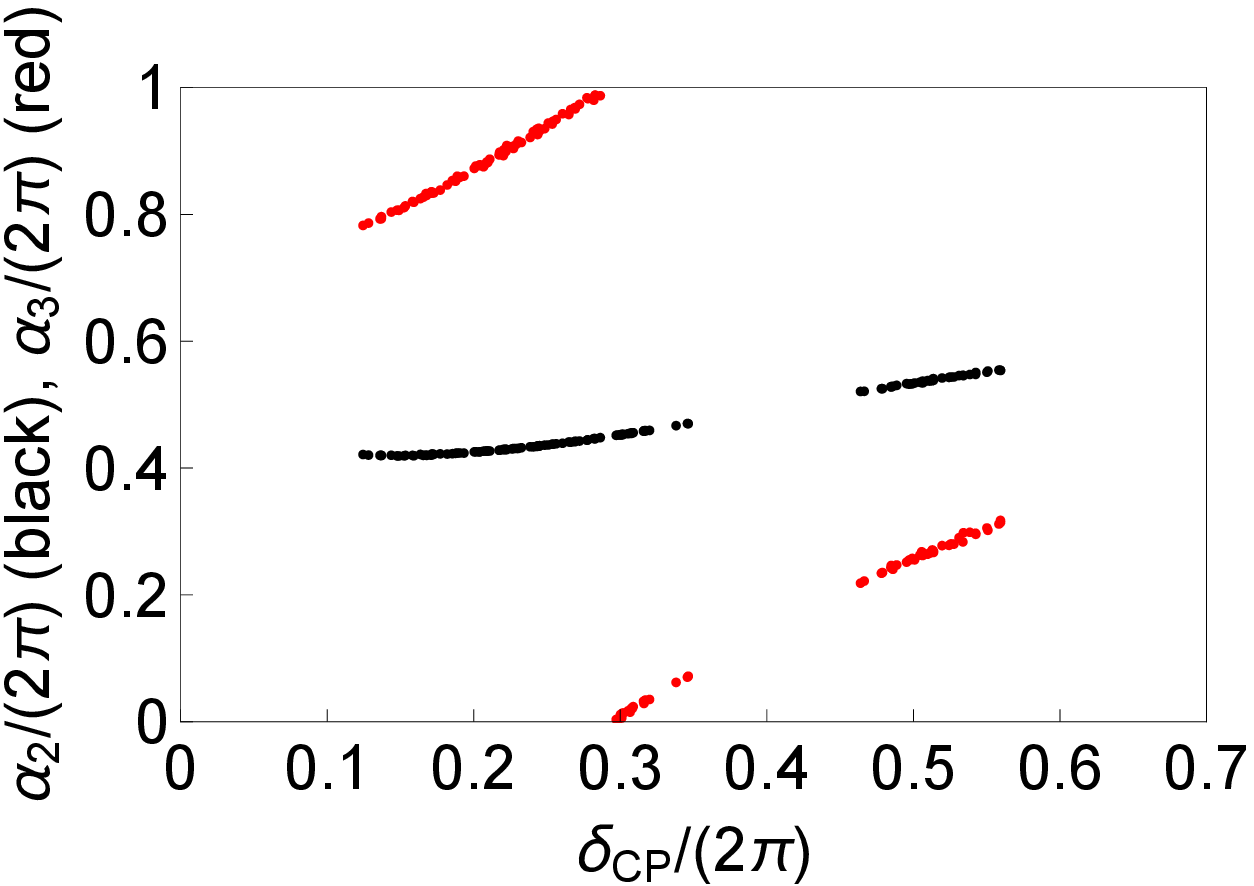}\\
\includegraphics[width=80mm]{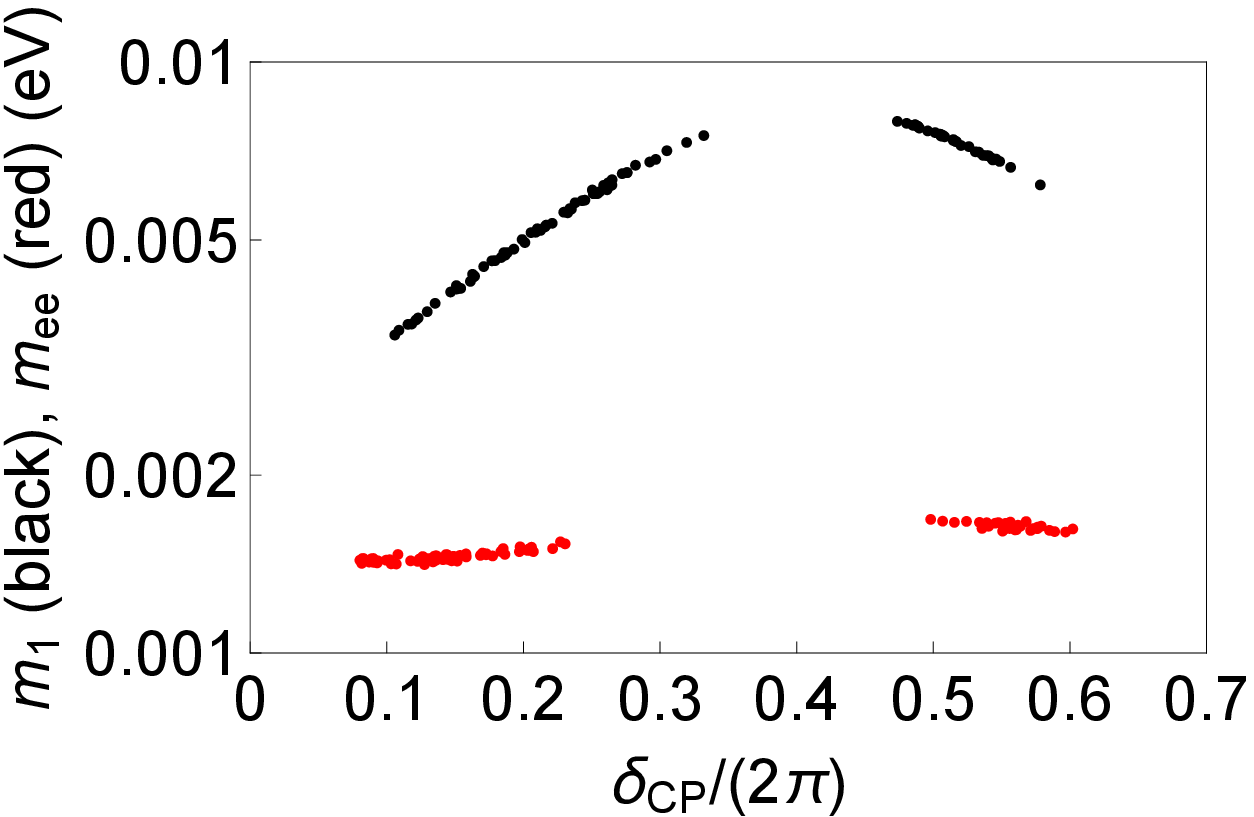}
\includegraphics[width=80mm]{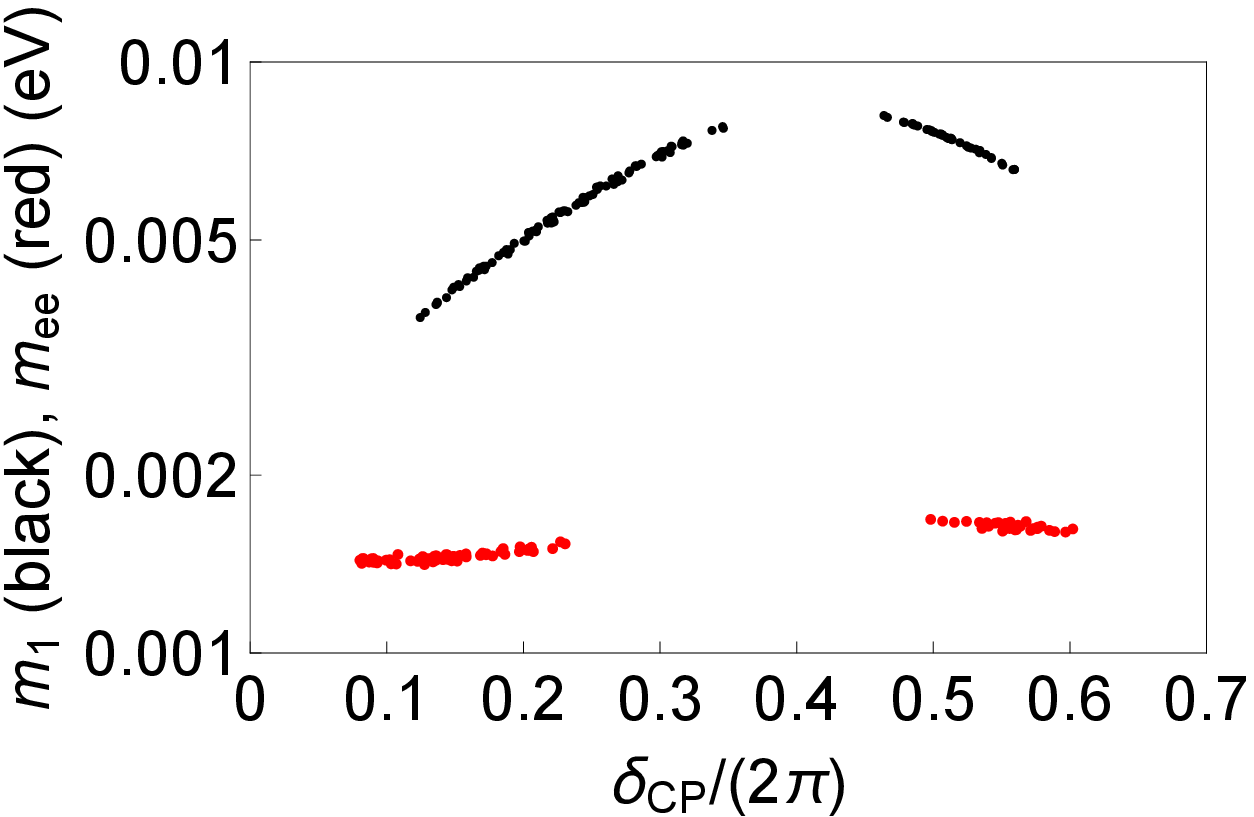}
\caption{Same as Figure~\ref{043047}, except that different input values as shown in Table~\ref{inputs} are used, 
 with the left plots corresponding to $y_b=5.293\times10^{-3}$
 and the right plots to $y_d=5.405\times10^{-3}$.
}
\label{ybpm}
\end{center}
\end{figure}

\item
We have varied the values of $\Delta m_{12}^2$ and $|\Delta m_{23}^2|$ within the $2\sigma$ experimental range,
 and found no significant change in the plots.

\item
In Figure~\ref{mu4fig}, we have used the values of $y_d,y_s,y_b$ and $\theta_{12}^{ckm},\theta_{13}^{ckm},\theta_{23}^{ckm},\delta_{km}$ at a different renormalization scale $\mu=10^4$~GeV (evaluated by taking $\mu_{EW}=M_Z$
and shown in Table~\ref{running})
 in Eqs.~(\ref{inputsrange},\ref{ckmselection}).
The neutrino oscillation parameters are identical with those for the right plots of Figure~\ref{043047}.
\begin{figure}[H]
\begin{center}
\includegraphics[width=80mm]{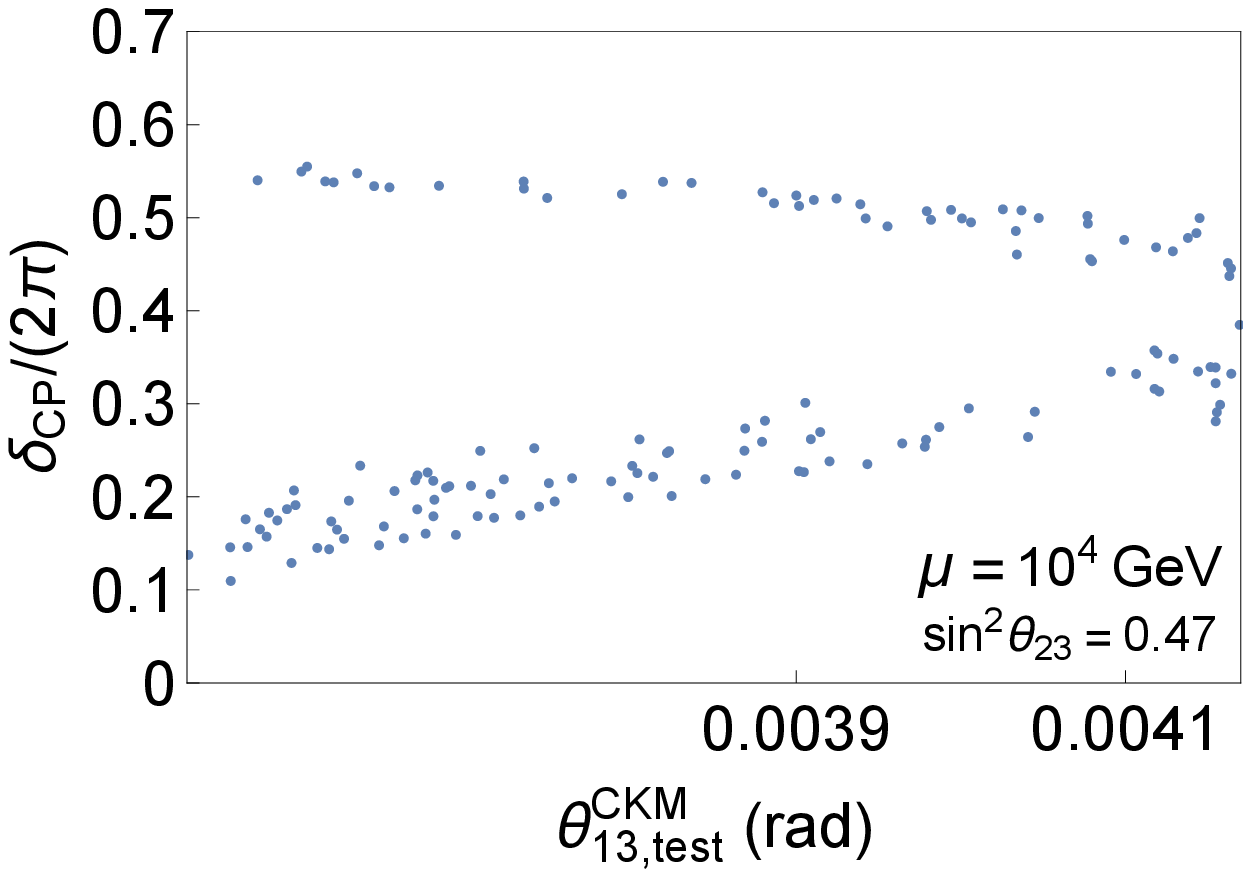}\\
\includegraphics[width=80mm]{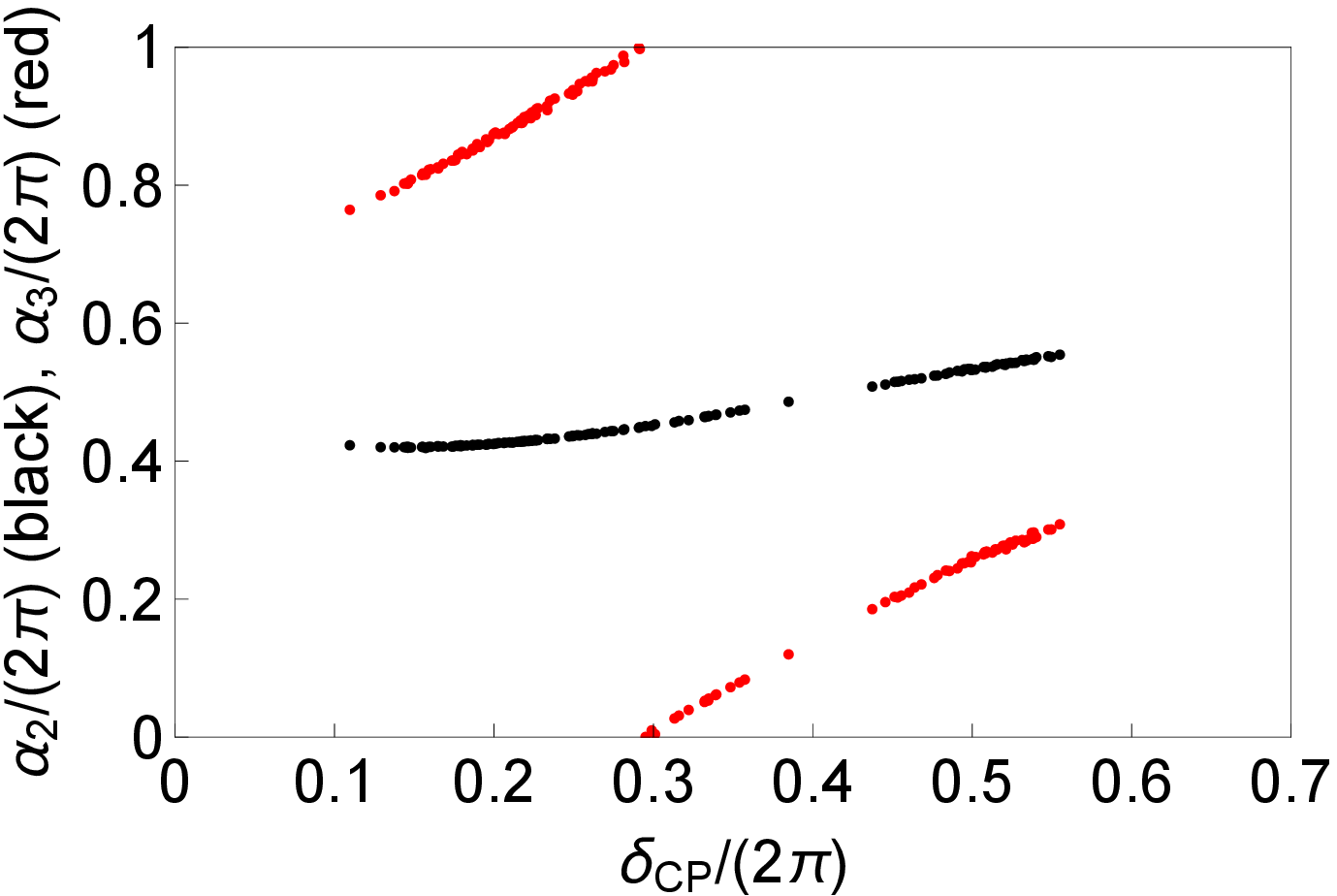}\\
\includegraphics[width=80mm]{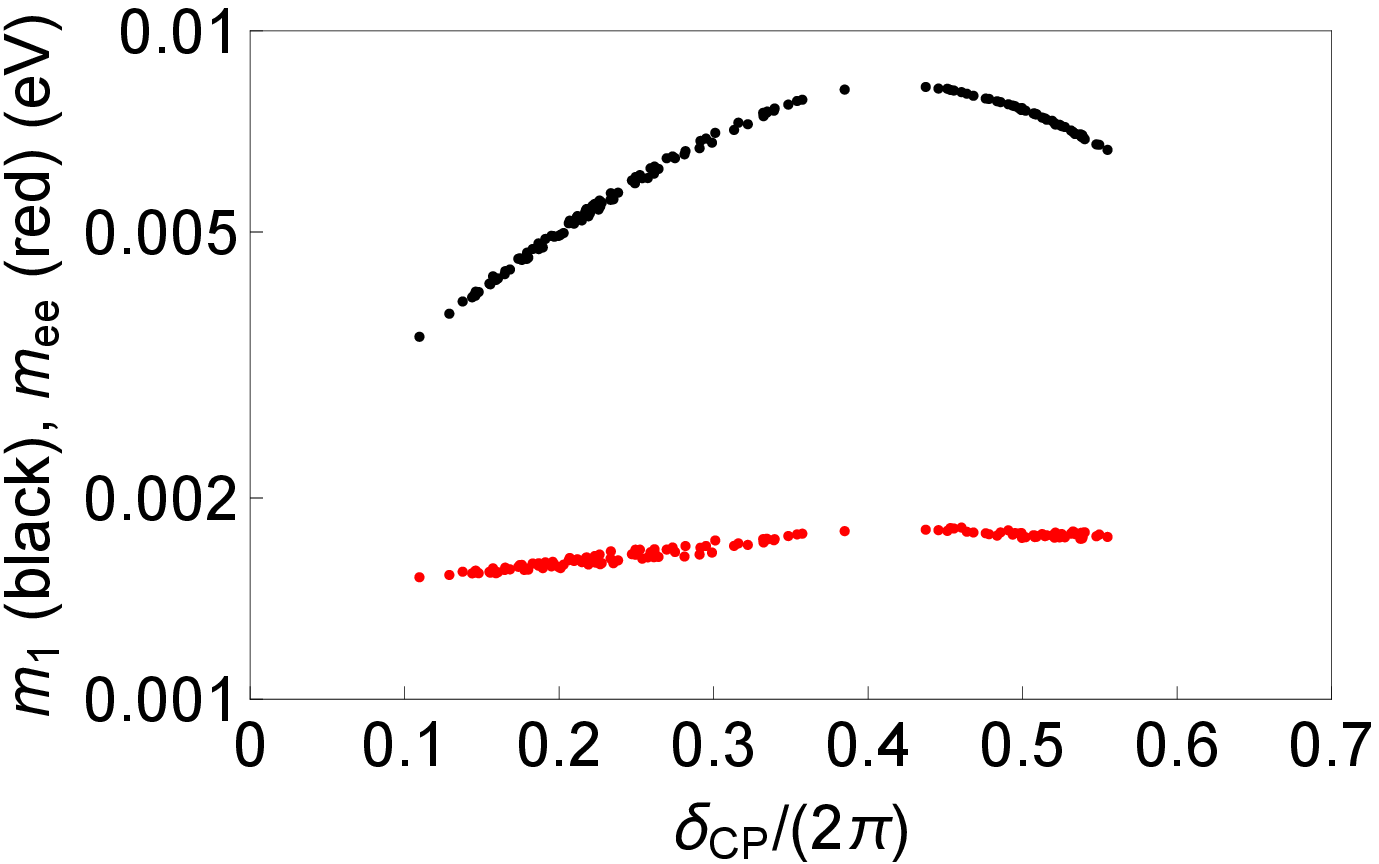}
\caption{Same as the right plots of Figure~\ref{043047}, except that the values of 
 $y_d,y_s,y_b$ and $\theta_{12}^{ckm},\theta_{13}^{ckm},\theta_{23}^{ckm},\delta_{km}$ at $\mu=10^4$~GeV,
 shown in Table~\ref{running}, are used in the analysis.
  The same values of neutrino oscillation parameters as the right plots of Figure~\ref{043047} are employed.
}
\label{mu4fig}
\end{center}
\end{figure}

 \item
To study the dependence of the results on the matching scale $\mu_{EW}$,
 we have used the values of $y_d,y_s,y_b$ and $\theta_{12}^{ckm},\theta_{13}^{ckm},\theta_{23}^{ckm},\delta_{km}$ at $\mu=10^{18}$~GeV
 evaluated by taking $\mu_{EW}=160$~GeV, shown in Table~\ref{running2}.
No significant difference is observed in the plots for $\mu_{EW}=M_Z$ and $\mu_{EW}=160$~GeV.

\end{itemize}

From the above results, the following observations are made:

\begin{itemize}

\item 
Only $Y_D \propto {\rm diag}(y_d,y_s,y_b)V_{CKM}^T$ Eq.~(\ref{yDyd}) can be consistent with the experimental data,
 and this is the case exclusively with the normal hierarchy of neutrino mass and for smaller values of $\sin^2\theta_{23}$
 in the current bound.

\item
In most cases, the value of $\delta_{CP}$ satisfying Eq.~(\ref{yDyd}) is in the range $\pi>\delta_{CP}>0$
 and hence is incompatible with the value hinted by the T2K collaboration, $\delta_{CP}\sim3\pi/2$~\cite{t2k}.
Nevertheless, for cases with $\sin^2\theta_{12}=0.333$, $y_d=5.36\times10^{-6}$ and $y_s=1.192\times10^{-4}$,
 there exist solutions with $\delta_{CP}\simeq 1.2\pi$.
Precise measurements of $\theta_{12}$ and improved evaluations
 of $s$ quark mass $m_s$ and $u$-$d$ quark mass ratio $m_u/m_d$
 (the average $(m_u+m_d)/2$ is well calculated compared to $m_u/m_d$)
 are required to falsify or corroborate our hypothesis.

\item
If we associate operators with opposite chiralities and consider the following different hypothesis,
\begin{align} 
Y_D &= z \, \begin{pmatrix} 
     1  & 0 & 0 \\
      0 & e^{i\phi_2} & 0 \\
      0 & 0 & e^{i\phi_3} \\
   \end{pmatrix} \begin{pmatrix} 
         y_d & 0 & 0\\
         0 & y_s & 0\\
         0 & 0 & y_b\\
      \end{pmatrix} V_{CKM}^\dagger
   \begin{pmatrix} 
     1  & 0 & 0 \\
      0 & e^{i\psi_2} & 0 \\
      0 & 0 & e^{i\psi_3} \\
   \end{pmatrix},
\label{yDyd2}
\end{align}
 then the sign of $\delta_{CP},\alpha_2,\alpha_3$ in Figures~\ref{043047},\,\ref{s12pm},\,\ref{s13pm},\,\ref{ydpm},\,\ref{yspm},\,\ref{ybpm} 
 is simply flipped.
This hypothesis is in good agreement with the T2K data on $\delta_{CP}$.

\item
The values of $\alpha_2,\alpha_3,m_1$ satisfying Eq.~(\ref{yDyd}) are strongly correlated with $\delta_{CP}$.
For $\delta_{CP}\simeq 1.2\pi$, $m_1$ is predicted to be about 0.005~eV,
 which may be tested in future cosmological observations (for forecasts, see, e.g., Ref.~\cite{forecast}).
$m_{ee}$ is suppressed below 0.002~eV due to cancellation of the active neutrino masses,
 and thus there is absolutely no chance to detect neutrinoless double $\beta$-decay in the near future~\cite{0nu}.

\item
The pattern of the correlation between $\delta_{CP}$ and $\theta_{13,test}^{ckm}$ is similar for
 the cases with $\mu=10^4$~GeV and $\mu=10^{18}$~GeV.
Since the running mixing angle $\theta_{13}^{ckm}$ depends linearly on the measured value of $|V_{ub}|$,
 we conclude that the correlation between $\delta_{CP}$ and $|V_{ub}|$ is nearly the same for $\mu=10^4$~GeV and $\mu=10^{18}$~GeV.
The plots for $\alpha_2,\alpha_3,m_1,m_{ee}$ are likewise the same for $\mu=10^4$~GeV and $\mu=10^{18}$~GeV.
Because the two distinctively different assumptions on the scale at which Eq.~(\ref{yDyd}) holds lead to similar results,
 we infer that our prediction is almost independent of the scale of Eq.~(\ref{yDyd}).

\item
We have confirmed that the above results are insensitive to the choice of the matching scale $\mu_{EW}$,
 which is reasonable because the ratio $y_d:y_s:y_b$ and the CKM mixing angles are intact with the change of $\mu_{EW}$, as read from Tables~\ref{running},\ref{running2}.
 
 \end{itemize}

To summarize, we have investigated whether the relation $Y_D \propto {\rm diag}(y_d,y_s,y_b)V_{CKM}^T$ or $Y_D \propto {\rm diag}(y_u,y_c,y_t)V_{CKM}^*$ (in the flavor basis where the charged lepton Yukawa coupling
 and right-handed neutrino Majorana mass are diagonal) can be consistent with
 the current experimental data on the quark masses, CKM mixing angles and phase, and neutrino mixing angles,
 hoping to unveil an implicit connection between the PMNS and CKM matrices.
We have found sets of values of $(\delta_{CP},\,\alpha_2,\,\alpha_3,\,m_1)$ that satisfy 
 $Y_D \propto {\rm diag}(y_d,y_s,y_b)V_{CKM}^T$ with the normal neutrino mass hierarchy,
  while there are no such sets for $Y_D \propto {\rm diag}(y_u,y_c,y_t)V_{CKM}^*$ and/or with the inverted hierarchy.
$\delta_{CP}$ is predicted to be in the range $1.2\pi\gtrsim\delta_{CP}>0$ and is hence in tension with
 the latest T2K data.
However, since the prediction crucially depends on neutrino mixing angles and $d,s$ quark masses,
 their future precise measurement or evaluation is necessary to draw any conclusion about our hypothesis.
We have made a prediction for $m_1$ that may be tested in future cosmological observations,
 whereas $m_{ee}$ is smaller than 0.002~eV and is far below the reach of near-future experiments.
\\

\section*{Acknowledgement}

The authors are indebted to Yukihiro Fujimoto (National Institute of Technology, Oita College),
 Hiroyuki Ishida (NCTS) and Yuya Yamaguchi for their contributions at the early stage of this work.
The authors are grateful to Mikhail Kalmykov (University of Hamburg) for informing us of important references. 
The authors would like to thank Tsuyoshi Nakaya (Kyoto University) for comments and encouragement.
This work is partially supported by Scientific Grants by the Ministry of Education, Culture, Sports, Science and Technology of Japan (Nos. 24540272, 26247038, 15H01037, 16H00871, and 16H02189).
\\

\end{document}